\documentclass{article}
\usepackage[english]{babel}
\usepackage[letterpaper,top=2cm,bottom=2cm,left=3cm,right=3cm,marginparwidth=1.75cm]{geometry}
\usepackage{amsmath,amssymb}
\usepackage{graphicx}
\usepackage[hidelinks]{hyperref}
\usepackage{setspace}
\usepackage{authblk}
\usepackage{booktabs}
\usepackage{amsthm}
\usepackage{natbib}
\usepackage[utf8]{inputenc}
\usepackage[T1]{fontenc}
\usepackage{titlesec}
\usepackage{array}              
\usepackage{rotating}
\usepackage{siunitx}            
\usepackage{threeparttable}     
\usepackage{multirow}           
\usepackage[font={small,bf}]{caption}
\usepackage{mathptmx}
\usepackage{algorithm}
\usepackage{algpseudocode}
\usepackage{tabularx,makecell}
\usepackage[table]{xcolor}
\usepackage{listings}
\usepackage{chngcntr}
\usepackage{placeins}  

\newtheorem{hyp}{Hypothesis}

\definecolor{backcolour}{gray}{0.95}
\definecolor{codegreen}{rgb}{0,0.6,0}
\definecolor{codegray}{rgb}{0.5,0.5,0.5}

\lstdefinestyle{mystyle}{
    backgroundcolor=\color{backcolour},   
    commentstyle=\color{codegreen},
    keywordstyle=\color{magenta},
    numberstyle=\tiny\color{codegray},
    stringstyle=\color{codegreen}, 
    basicstyle=\ttfamily\footnotesize,
    breakatwhitespace=false,         
    breaklines=true,                 
    captionpos=b,                    
    keepspaces=true,                 
    numbers=left,                    
    numbersep=5pt,                  
    showspaces=false,                
    showstringspaces=false,
    showtabs=false,                  
    tabsize=2,
    frame=single,
    framerule=0.5pt
}
\title{Validating Generative Agent-Based Models for Logistics and Supply Chain Management Research}
\author{Vincent E. Castillo, PhD}
\affil{Fisher College of Business, The Ohio State University}
\date{}

\begin{document}
\maketitle

\begin{abstract}
\noindent Generative Agent-Based Models (GABMs) powered by large language models (LLMs) offer promising potential for empirical logistics and supply chain management (LSCM) research by enabling realistic simulation of complex human behaviors. Unlike traditional agent-based models, GABMs generate human-like responses through natural language reasoning, which creates potential for new perspectives on emergent LSCM phenomena. However, the validity of LLMs as proxies for human behavior in LSCM simulations is unknown. This study evaluates LLM equivalence of human behavior through a controlled experiment examining dyadic customer-worker engagements in food delivery scenarios. I test six state-of-the-art LLMs against 957 human participants (477 dyads) using a moderated mediation design. This study reveals a need to validate GABMs on two levels: (1) human equivalence testing, and (2) decision process validation. Results reveal GABMs can effectively simulate human behaviors in LSCM; however, an equivalence-versus-process paradox emerges. While a series of Two One-Sided Tests (TOST) for equivalence reveals some LLMs demonstrate surface-level equivalence to humans, structural equation modeling (SEM) reveals artificial decision processes not present in human participants for some LLMs. These findings show GABMs as a potentially viable methodological instrument in LSCM with proper validation checks. The dual-validation framework also provides LSCM researchers with a guide to rigorous GABM development. For practitioners, this study offers evidence-based assessment for LLM selection for operational tasks.\\

\noindent\textbf{Keywords:} large language models, artificial intelligence, generative agent based model, platform operations, last mile logistics
\end{abstract}

\doublespacing

\section{Introduction}
Understanding the system-level implications of complex human behaviors is an important pursuit in Logistics and Supply Chain Management (LSCM) research. Particularly interesting behavioral phenomena reside in dyadic LSCM interactions (e.g., customer-worker engagements, buyer-supplier negotiations, or shipper-carrier relationships), which fundamentally shape operational outcomes and competitive advantage \citep{nyaga_power_2013, ribbink_impact_2014, brito_power_2017, belhadi_behavioral_2021}. These dyadic relationships involve psychological processes, social dynamics, and power asymmetries, which make the resulting emergent behaviors and system-level impacts difficult to measure empirically \citep{kenny_dyadic_2006}.

Agent-based models (ABMs) represent one approach to capture the system-level consequences of these complex behavioral dynamics, whereby researchers simulate interactions between multiple system constituents to observe emergent phenomena \citep{abar_agent_2017, bonabeau_agent-based_2002, kasaie_guidelines_2015, macal_everything_2016}. ABM applications in LSCM include supply chain resilience \citep{zhao_modelling_2019}, last mile delivery \citep{castillo_designing_2022}, managing research and development projects \citep{chandrasekaran_managing_2016}, and humanitarian logistics \citep{altay_information_2014}. However, ABMs approximate human behavior with mathematical or programmatic rules that, to ensure mathematical tractability, typically include simplifying assumptions that may not reflect empirical reality \citep{law_simulation_2015, evers_systems_2012}. Furthermore, LSCM dyads are marked by complex human cognitive processes, psychological biases, and contextual nuances that rule-based systems cannot fully encode or capture. For this reason, current ABM approaches often result in analyses that focus on particular rather than generalized system behavior \citep{mcgrath_dilemmatics_1981}.

Pursuing deeper understanding of complex human behavior, including dyadic interactions and their system-level implications, is the locus of behavioral LSCM research. Lab and field behavioral experiments provide the foundation for understanding the psychological and cognitive processes of how humans make LSCM decisions \citep{carter_experiments_2024, eckerd_making_2021, gao_field_2023, lee_running_2018, lonati_doing_2018, rungtusanatham_vignette_2011, ta_designing_2025}. While behavioral experiments allow for deep understanding of psychological processes, especially in dyads, assessing the emergent implications is typically impossible due to inability to measure how the real-world LSCM system responds to experimental manipulations \citep{mcgrath_dilemmatics_1981}.

To bridge this gap between psychological realism and system-level analysis, Generative Agent-Based Models (GABMs) powered by Large Language Models (LLMs) represent the next evolution in the ABM tradition by enabling agents to generate human-like responses through natural language reasoning rather than pre-programmed behavioral rules \citep{xie_can_2024, ghaffarzadegan_generative_2024, wu_smart_2023, park_generative_2023, vezhnevets_generative_2023}. LLMs have been shown to emulate surface-level human behaviors convincingly \citep{xie_can_2024} and reliably predict human behaviors \citep{binz_foundation_2025}. This natural language capability allows agents to process contextual nuances, social cues, and situational dynamics that rule-based ABMs cannot, which appears to enable more realistic simulation of human behaviors. GABMs purport to combine the power of LLMs to approximate human behavior with the system simulation capabilities of ABMs. This combination may allow researchers to explore more realistic systems and simulate theoretical mechanisms that create emergent outcomes \citep{miller_agent-based_2015, tornberg_abstractions_2019}.

However, the validity of LLMs for human decision-making in LSCM specifically, either as customers, workers, or managers, for example, is unknown. In fact, no systematic frameworks exist to assess whether LLMs reliably simulate dyadic human behavior \citep{larooij_large_2025}, let alone in LSCM. However, the ABM tradition (and simulation in general) has numerous frameworks and guidelines for simulation model development that can inform rigorous GABM development \citep[e.g.][]{bowersox_simulation_1989, rand_agent-based_2011, gurcan_generic_2013, sargent_verification_2013, law_simulation_1982, shafer_empirical_2004, bonabeau_agent-based_2002, kleijnen_verification_1995}. This research draws upon historical simulation research to develop a GABM validation framework by incorporating criteria explicitly designed to validate the behavior of generative agents in LSCM settings. Specifically, this research addresses two questions: \textit{(1) Can GABMs powered by LLMs satisfactorily emulate surface-level human behaviors in LSCM dyadic interactions? (2) Do these generative agents demonstrate decision-making processes comparable to humans?}

To explore how closely LLM and human outcomes align at multiple levels, this research applies the GABM framework to a case study of customer-worker dyads in online food delivery platforms, such as UberEATS or DoorDash. Food delivery represents a particularly important dyadic context to study given its rapid growth, high-frequency social interactions, and reliance on technology-mediated coordination between customers and gig workers \citep{ta_designing_2018, castillo_crowdsourcing_2018, carbone_rise_2017, allon_impact_2023, dayarian_crowdshipping_2020, miao_effects_2023}. Unlike buyer-supplier dyads, which are characterized by long-term relationships and contractual governance, or shipper-carrier relationships involving established networks, food delivery dyads feature brief, service-focused encounters with high emotional stakes and immediate feedback mechanisms \citep{wang_effects_2025, simoni_potential_2020, saunders_improving_2025}. Critically, these platforms operate as multi-sided markets where workers function not merely as service providers, but as essential stakeholders whose satisfaction and retention are fundamental for platform viability. As a result, these interactions mutually influence both customer and worker satisfaction, with both sides' experiences having implications for maintaining sufficient network effects for ongoing platform operations \citep{wang_effects_2025}. While food delivery dyads do not represent all LSCM dyadic relationships, they provide a controlled context for developing a GABM validation methodology that can be adapted to other dyadic contexts.

The proposed GABM development framework provides a dual-validation approach offering both surface-level equivalence testing and decision-process validation, allowing researchers to select the appropriate validation level(s) for their specific research context. Equivalence testing assesses how closely LLM outputs correspond to human outputs, while decision-process validation examines the decision-making processes of LLMs relative to humans. The study systematically evaluates six state-of-the-art LLMs: OpenAI's \textit{GPT-4o} and \textit{GPT-4.1}, Anthropic's \textit{Claude Sonnet 3.5} and \textit{Claude Sonnet 4}, and Mistral AI's \textit{Large 2} and \textit{Medium 3} models, which were selected for their advanced capabilities and widespread industry adoption. This model portfolio also allows for within-family qualitative comparisons of two recent generations (2024 and 2025) of LLMs. The six LLMs are compared to 957 human participants organized into 477 food delivery customer-worker dyads. The analysis begins with a series of Two One-Sided Tests (TOST) to assess LLM equivalence of surface-level human behaviors \citep{lakens_equivalence_2018, lakens_equivalence_2017, rainey_arguing_2014}. Then, a structural equation model (SEM) is estimated with bootstrap analysis of a moderated mediation design to assess decision-process validity in the LSCM dyad \citep{shrout_mediation_2002, efron_introduction_1994, hayes_introduction_2017}. 

The findings reveal an interesting tension: surface-level behavioral equivalence does not guarantee that LLMs replicate human decision-making processes in GABM applications. For example, \textit{GPT-4o} was the only LLM to demonstrate surface-level equivalence to humans on three behavioral measures, yet performed the worst in process-level validation with only 5/10 pathway matches to human decision-making patterns. Conversely, four AI models—\textit{GPT-4.1}, \textit{Sonnet 3.5}, \textit{Sonnet 4}, and \textit{Mistral Medium 3}—achieved the highest process-level fidelity (8/10 pathway matches) despite demonstrating weaker surface-level equivalence, with most achieving equivalence on only one behavioral measure. Surprisingly, \textit{Mistral Large 2}, while being the only AI to correctly replicate both human indirect effects, ranked in the middle for overall process validation (6/10 matches) due to direct pathway deviations. Additionally, ``state-of-the-art'' (2025) models were less likely than earlier generation (2024) models to show surface-level equivalence, which raises questions model improvement trajectories and alignment with human behavior. These results provide preliminary evidence that LLMs may serve as viable human behavioral proxies in simulations, contingent upon systematic validation tied to the specific modeling context. Importantly, this study does not suggest that LLMs should replace humans in behavioral research; rather, it demonstrates that properly validated LLMs may function as behavioral proxies in simulation environments, just as traditional simulations serve as proxies for real LSCM systems.

The research makes three theoretical contributions to the LSCM literature. First, a dual-validation framework is developed for using GABMs in LSCM research. This framework enables researchers to choose between surface-level behavioral validation, process-level decision validation, or both, depending on the specific requirements of the phenomenon being studied. Second, the GABM case study demonstrates that food delivery platform satisfaction operates as a dyadic phenomenon where both customer and worker experiences are simultaneously shaped by service quality, with the timing of tip information presentation aligning dyadic satisfaction. The findings extend the concept of consumer-centric platform design into dyad-centric design by establishing that worker satisfaction should be treated as equally important to platform operations as customer satisfaction. Third, the study develops a future research agenda where GABMs could be applied in LSCM. Managerially, the research provides empirical guidelines for selecting and validating LLMs in operational contexts, which promotes responsible GenAI adoption in LSCM.

\section{Literature Review}

\subsection{Agent-Based Modeling in Logistics and Supply Chain Management}

Agent-based modeling is a simulation approach to understanding complex system behavior \citep{bonabeau_agent-based_2002, hall_visualizing_2016}. Collections of autonomous entities called ``agents'' interact in a simulated environment with the goal of producing emergent system-level phenomena whose causality is attributable to the complexity of the constituent elements' structural arrangement and interactions \citep{bonabeau_agent-based_2002, abar_agent_2017, rand_agent-based_2011, smith_simulating_2018, macal_everything_2016, kasaie_guidelines_2015}. This bottom-up methodology distinguishes ABM from top-down approaches like discrete event simulation, which models systems as a series of processes \citep{law_simulation_2015}. The ABM perspective enables researchers to uncover generative mechanisms and emergent behaviors that may not be apparent in aggregate-level analysis \citep{tornberg_abstractions_2019}. The underlying philosophy posits that by simulating micro-level interactions, researchers can gain deeper understanding of macro-level outcomes \citep{miller_agent-based_2015}, which is a particularly relevant principle for LSCM systems where individual stakeholder behaviors aggregate to create complex supply chain dynamics.

ABM studies in LSCM have examined foundational operational challenges including supply chain risk and resilience \citep{zhao_modelling_2019, chen_supply_2022, basole_supply_2014}, healthcare operations \citep{gibbons_designing_2009, ayer_prioritizing_2019}, supply chain networks as complex adaptive systems \citep{choi_supply_2001, giannoccaro_impact_2018, li_out_2021}, urban freight systems and last mile delivery \citep{ambra_you_2021, castillo_designing_2022, castillo_hybrid_2021}, and humanitarian logistics \citep{altay_information_2014}. Each of these studies models humans or organizations as decision-making agents that interact with other agents where the system-level performance is theorized to be the outcome of the micro-level interactions. These applications demonstrate ABM's capacity to model distributed decision-making processes and capture the emergent macro-level outcomes of multiple stakeholder interactions that characterize modern supply chains.

Behavioral representation in traditional ABMs relies primarily on programmed rules, decision trees, and algorithm-following agents that follow predetermined logic structures \citep{rand_agent-based_2011, smith_simulating_2018}. These mathematical formulations specify agent actions based on environmental conditions, resource constraints, and interaction protocols, with agents exhibiting behaviors that range from simple scripted responses to more sophisticated adaptive mechanisms \citep{hall_visualizing_2016, macal_everything_2016}. At the simpler end of this spectrum, Schelling's \citeyearpar{schelling_models_1969} segregation model programmed agents with basic preference thresholds that triggered relocation decisions through straightforward if-then logic. More sophisticated implementations, such as Castillo et al.'s \citeyearpar{castillo_designing_2022} adaptive last-mile delivery agents, employ dynamic learning mechanisms based on exponential smoothing functions that allow agent behavior to evolve based on accumulated interaction experiences with customers. Regardless of their complexity level, these rule-based approaches share the advantage of enabling tractable mathematical analysis and computational efficiency, making them well-suited for large-scale system modeling while maintaining the fundamental limitation of approximating human behavior through predetermined algorithmic structures.

However, significant limitations arise when attempting to model authentic dyadic interactions using rule-based approaches \citep{siebers_discrete-event_2010}. Customer-worker relationships, buyer-supplier negotiations, and shipper-carrier communications involve contextual nuances, emotional responses, and adaptive communication patterns that resist mathematical specification. Pre-programmed behavioral rules struggle to capture the psychological processes, social dynamics, and situational adaptability that characterize human decision-making in LSCM contexts. While ABM has proven valuable for understanding system-level phenomena in LSCM, the field collectively acknowledges that pre-programmed behavioral rules constrain the authenticity of human representation \citep{evers_systems_2012}, particularly in complex dyadic interactions where contextual nuances and adaptive responses are critical.

\subsection{Human Behavioral Modeling in LSCM Dyadic Interactions}
Dyadic relationships form the fundamental building blocks of LSCM processes, with customer-worker interactions representing a contemporary and particularly salient  category for delivery platforms. Research on service operations and dyadic relationships has established that the quality of interpersonal interactions significantly influences customer satisfaction, loyalty, and firm performance outcomes in last mile logistics contexts \citep{masorgo_youre_2023, zhou_platform_2022, saunders_improving_2025}. These dyadic encounters involve complex psychological processes including expectation formation, service quality perception, and satisfaction evaluation that directly impact operational metrics and strategic outcomes.

Food delivery platforms exemplify the importance of dyadic interactions in contemporary LSCM contexts with novel forms of platform-mediated interactions between customers and gig workers \citep{zhao_modelling_2019, ta_designing_2025, ta_crowdsourced_2023, ta_designing_2018, dayarian_crowdshipping_2020, simoni_potential_2020}. These interactions demonstrate how service encounter quality affects both customer and worker satisfaction. Mutual influence patterns create interdependent outcomes that determine platform economic sustainability \citep{wang_effects_2025}. The technology-mediated nature of these relationships introduces additional complexity through information asymmetries, rating and tipping mechanisms, and algorithmic matching processes that influence behavioral outcomes \citep{zhou_platform_2022, wang_scalability_2022}.

Trust and relationship quality research in LSCM contexts has identified communication patterns, reliability perceptions, and relationship development processes as critical factors in LSCM dyads \citep{zhou_platform_2022}. Psychological processes in LSCM dyads encompass emotional responses, cognitive biases, expectation formation, and social influence mechanisms that shape decision-making and satisfaction outcomes for customers and workers \citep{ta_crowdsourced_2023, saunders_improving_2025}. Customer disappointment following service failures, worker frustration with unrealistic demands, and satisfaction derived from exceeding expectations all represent psychological phenomena that influence operational outcomes but resist direct measurement and modeling in traditional ABM.

Current methodological approaches for studying dyadic processes such as surveys or lab experiments face significant constraints. Surveys capture only stated intentions rather than observed behavior, and lab experiments may not reflect real-world complexity assessing private dyadic interactions \citep{kenny_dyadic_2006, rungtusanatham_vignette_2011, ta_designing_2025}. These methodological challenges collectively have a fundamental limitation: while researchers can study individual psychological processes in controlled settings, they cannot easily observe how these micro-level processes scale to create emergent system-level phenomena in operational contexts \citep{mcgrath_dilemmatics_1981}.

LSCM behavioral research has successfully identified key psychological mechanisms driving dyadic interactions, yet methodological constraints prevent researchers from observing how these micro-level processes scale to create emergent system-level phenomena. This is a fundamental limitation that restricts both theoretical development and practical application.

\subsection{Large Language Models and Generative AI for Behavioral Simulation}

Large Language Models (LLMs) have demonstrated capabilities for simulating human-like reasoning, decision-making, and communication across diverse contexts, which creates new possibilities for behavioral modeling in LSCM research \citep{shalpegin_generative_2025}. These models exhibit emergent behaviors that appear to replicate human cognitive processes including contextual reasoning, emotional response generation, and adaptive communication strategies. Unlike traditional rule-based ABMs, LLMs generate responses through learned patterns from extensive training data, enabling more nuanced and contextually-grounded behavioral simulation.

LLM applications in LSCM have primarily focused on customer service automation, supplier communication optimization, and stakeholder interaction management \citep{spring_how_2022, simchi-levi_large_2025}. These applications demonstrate LLMs' capacity to process unstructured communication, interpret contextual cues, and generate appropriate responses in business contexts. However, the current literature on LLMs in LSCM is nascent, with applications emphasizing efficiency and automation in industry rather than authentic behavioral simulation, leaving questions about psychological fidelity largely unexplored.

Evidence from psychology, behavioral economics, and social science disciplines provides mixed signals regarding LLM behavioral fidelity \citep{shalpegin_generative_2025}. Some studies demonstrate impressive alignment between LLM responses and human behavioral patterns \citep{binz_foundation_2025, liu_toward_2025, park_generative_2023, vezhnevets_generative_2023}, while others reveal systematic biases, inconsistencies, and artificial reasoning patterns that diverge from authentic human cognition. These findings highlight the importance of context-specific validation rather than assuming general behavioral equivalence across all domains.

The generative capacity of AI allows a fundamental departure from traditional behavioral modeling approaches \citep{junprung_exploring_2023, shalpegin_generative_2025}. Text generation capabilities enable agents to produce contextually grounded responses, engage in multi-turn interactions, and adapt communication strategies based on situational demands. This flexibility offers significant advantages over pre-programmed behavioral rules, potentially enabling more realistic simulation of complex stakeholder interactions in logistics contexts.

However, significant limitations and concerns persist regarding LLM reliability, bias, hallucination tendencies, and behavioral inconsistency in applied contexts \citep{junprung_exploring_2023, shalpegin_generative_2025}. Models may produce plausible-sounding responses that reflect training data biases rather than authentic human behavioral patterns. The opacity of neural network decision-making processes creates challenges for understanding whether LLMs achieve behavioral equivalence through authentic reasoning or superficial pattern matching, although there is a similar question for human subjects as well. Further, while LLMs demonstrate promising capabilities for human behavioral simulation across multiple disciplines, their validity for LSCM-specific decision-making contexts remains unexplored, and no systematic frameworks exist for assessing whether these models reliably capture the psychological processes that drive dyadic LSCM behaviors.

\subsection{Simulation Validation Frameworks}

Concerns about LLM behavioral fidelity stem from training data and the black-box nature of the neural network transformer architecture that governs the text generation process, but also the stochastic nature of generative AI models. While the first two concerns are ongoing areas of inquiry in AI science, the operations research (OR) discipline provides established guidance for validating stochastic systems. Simulation validation is essential for ensuring model reliability and appropriateness for intended applications \citep{sargent_verification_2013, shafer_empirical_2004, law_simulation_1982, rand_agent-based_2011}. This suggests that historical simulation validation approaches can be adapted to account for LLM stochasticity in GABMs.

Traditional simulation validation encompasses multiple validation types including conceptual model validity, computerized model verification, operational validity, and data validity \citep{sargent_verification_2013}. ABM-specific frameworks extend this approach through hierarchical validation at multiple levels of abstraction, recognizing that models may perform adequately at one level while failing at others \citep{rand_agent-based_2011, macal_everything_2016}. Micro-level validation ensures individual agent behaviors and interaction patterns are appropriate, while macro-level validation confirms system-level outcomes are suitable approximations for the study's intended purpose.

Behavioral validation presents unique challenges that distinguish it from traditional simulation validation \citep{lonati_doing_2018, ta_reconceptualizing_2025}. Rather than focusing solely on statistical similarity between model outputs and empirical data, behavioral validation requires assessing whether simulated scenarios authentically match real world phenomena and subsequent human behaviors through appropriate psychological processes \citep{eckerd_making_2021, carter_experiments_2024}. This introduces a fundamental distinction between outcome and process validation: separating validation of surface-level outputs from validation of underlying decision mechanisms is essential to determining whether LLMs can provide suitable proxies for human behavior in GABMs.

A model might produce statistically equivalent outcomes to human behavior while employing completely artificial decision processes, or conversely, demonstrate authentic reasoning patterns yet failing to match empirical outcome distributions. This process-versus-outcome distinction is analogous to macro/micro level validation in traditional simulation and is critical for GABMs, particularly when models are used for policy design or strategic decision-making where understanding behavioral mechanisms matters as much as predicting outcomes.

Simulation and OR disciplines provide robust frameworks for model validation, but these approaches were developed for mathematical and rule-based models. The field has yet to develop validation methodologies specifically designed for LLM-based or ``generative'' agents that produce behavior through natural language reasoning. This creates a critical gap in the ability to assess AI-powered behavioral simulations.

\subsection{Generative Agent-Based Models: An Emerging Paradigm}
GABMs represent a fundamental evolution from traditional ABM methodology through integration of natural language reasoning capabilities that enable agents to generate contextually appropriate behaviors without explicit rule programming. Recent advancements in AI have led to the emergence of GABMs that combine the strengths of traditional ABMs with the novel capabilities of LLMs \citep{park_generative_2023, vezhnevets_generative_2023, gao_large_2023}. These developments open new possibilities for simulating complex human behaviors and social interactions with enhanced realism and flexibility.

Early GABM studies have demonstrated the potential for creating believable human behavioral simulation in controlled environments. Park et al. \citeyearpar{park_generative_2023} introduced a comprehensive architecture for generative agents capable of performing daily activities, forming opinions, initiating conversations, and planning for future actions within sandbox environments. Their work established key principles for how LLMs can be integrated into agent frameworks to produce dynamic and adaptive simulations of social systems, thereby providing a template for subsequent GABM development.

Additional GABM applications across diverse domains illustrate both the potential and current limitations of this emerging paradigm. Gao et al. \citeyearpar{gao_large_2023} provide a comprehensive survey of LLM applications in ABM and simulation, highlighting capabilities across cyber, physical, social, and hybrid environments. Ghaffarzadegan et al. \citeyearpar{ghaffarzadegan_generative_2024} demonstrate GABM applications through a case study of social norm diffusion in organizational contexts, showing how GABMs can incorporate realistic human reasoning and decision-making into dynamic models of social systems.

Dyadic interaction capabilities represent a particularly relevant application area for LSCM contexts. Junprung \citeyearpar{junprung_exploring_2023} investigated LLM-powered simulations of human interactions through two-agent negotiations and multi-agent collaborative scenarios, demonstrating the potential for sophisticated and accurate simulations of human-driven dyadic encounters. These applications directly parallel the customer-worker, buyer-supplier, and shipper-carrier dyads that characterize LSCM operations, suggesting GABMs' relevance for logistics research applications.

Technical implementation approaches have evolved to address the complexity of integrating LLMs into agent-based frameworks. Wu et al. \citeyearpar{wu_smart_2023} proposed Smart Agent-Based Modeling (SABM) frameworks that leverage LLMs to simulate real-world scenarios with increased nuance and realism, presenting case studies demonstrating effectiveness in modeling complex systems. Vezhnevets et al. \citeyearpar{vezhnevets_generative_2023} developed Concordia, a comprehensive library for constructing GABMs that introduces ``Game Master'' agents responsible for simulating environments where generative agents interact, enabling language-mediated simulations in both physical and digital contexts.

Advanced GABM architectures have begun addressing psychological validity concerns through more sophisticated cognitive modeling approaches. Mitsopoulos et al. \citeyearpar{mitsopoulos_psychologically-valid_2023} introduced Psychologically-Valid Generative Agents that combine cognitive architectures with LLMs to create more realistic behavioral simulations, integrating data-driven decision-making with human-like linguistic generation for applications in public health, group dynamics, and financial markets. Xiao et al. \citeyearpar{xiao_simulating_2023} proposed the Generative Agent-Based Simulation System (GABSS) specifically designed to emulate human cognition, memory, and decision-making frameworks within virtual social systems, highlighting capabilities for personalized agent customization and natural language processing integration.

Despite these promising developments, significant methodological challenges remain unresolved, especially in LSCM contexts. Current GABM applications often lack systematic validation against human behavioral baselines, raising questions about the authenticity of simulated behaviors versus superficial pattern matching. Issues of bias mitigation, results interpretation, and validation methodology represent critical gaps that limit confidence in GABM deployment for empirical research or policy applications \citep{shalpegin_generative_2025}.

GABMs represent a promising evolution of traditional ABM methodology with demonstrated capabilities for dyadic behavioral simulation relevant to LSCM contexts, yet their application to logistics research remains largely theoretical. The field lacks both empirical evidence of GABM effectiveness in LSCM contexts and systematic approaches for implementing, validating, and selecting appropriate models for LSCM research applications. 

\section{A Multi-Level Validation Framework for Generative Agent-Based Models}

Traditional simulation verification and validation methodology provides established techniques for ensuring model reliability and appropriateness, but requires adaptation for the unique challenges presented by generative agents \citep{sargent_verification_2013}. Both discrete-event simulation and agent-based modeling traditions emphasize multi-level validation requirements, recognizing that rigorous validation must occur at individual and system levels to demonstrate methodological adequacy. GABMs present novel validation challenges because AI agents can potentially achieve behavioral equivalence while employing completely artificial underlying decision processes.

Surface-level validation assesses behavioral equivalence of generative agents with humans, focusing on the ``what'' of decision-making rather than underlying processes. The central question becomes whether AI agents make the same decisions as humans when faced with identical scenarios. The comparisons can be assessed through statistical equivalence testing approaches such as Two One-Sided Tests (TOST) that determine whether behavioral differences fall within acceptable bounds \citep{lakens_equivalence_2017, lakens_equivalence_2018, rainey_arguing_2014}.

Decision-process validation examines whether AI agents follow similar cognitive pathways and exhibit similar psychological mechanisms as humans \citep{smith_simulating_2018}. This validation level focuses on the ``how'' and ``why'' of decision-making \citep{stank_new_2017, craighead_goldilocks_2016, craighead_using_2024}, examining mental steps, situational interpretations, and causal reasoning patterns that lead to system outcomes. Process validation can be assessed through methods such as Structural Equation Modeling (SEM) to test whether generative agents demonstrate the same causal pathways observed in human decision-making \citep{shrout_mediation_2002, efron_introduction_1994, hayes_introduction_2017}.

The need for multi-level validation stems from the core principle that single-level validation is insufficient for rigorous simulation methodology \citep{law_simulation_1982, rand_agent-based_2011}. GABMs face the particular challenge that AI models can pass surface-level validation (output equivalence) while failing process-level validation (authentic decision processes), or conversely, demonstrate authentic reasoning patterns while producing outputs that differ statistically from human baselines. These two validation levels are complementary yet distinct in that success at one level does not guarantee success at the other. This dual possibility necessitates systematic assessment at both levels to ensure appropriate model selection and deployment.

This multi-level validation framework extends established LSCM simulation methodology to address the unique challenges of validating AI-powered behavioral agents in LSCM contexts \citep{sargent_verification_2013, shalpegin_generative_2025}. While demonstrated through statistical equivalence testing and structural equation modeling in this study, the conceptual framework remains methodology-independent, thus allowing researchers to employ different analytical approaches depending on research context and validation objectives. The framework provides systematic standards for GABM validation that align with simulation science best practices while addressing the novel challenges of generative behavioral modeling.

\section{GABM Case Study -- On-demand Food Delivery Platforms} \label{sec:caseStudy}
\subsection{Multi-Sided Platform Design in Food Delivery}

Multi-sided platforms represent a distinct class of business models where different agent populations interact through an intermediary, and the decisions of one agent set affect the outcomes of another \citep{rysman_economics_2009}. Food delivery platforms are one such example where the need to simultaneously attract and retain both customers and workers is essential \citep{saunders_improving_2025}. In fact, the economic sustainability of platforms depends on the ability to facilitate interactions between agent populations while maintaining satisfaction on both sides of the market.

Tipping plays a fundamental role in these food delivery customer-worker interactions. While traditionally serving as a reward for exceptional service \citep{castillo_designing_2022}, tipping has evolved into a foundational component of platform workers' total remuneration \citep{zhang_allow_2024, chen_value_2019}. This transformation creates potential for negative feedback loops: when delivery workers experience dissatisfaction due to misaligned tipping expectations, service quality may decline for subsequent deliveries \citep{yang_risks_2025}. Additionally, platforms typically allow customers to adjust tip amounts after service delivery, creating the possibility of ``tip baiting'' where customers initially leave high tips then remove them after receiving delivery \citep{obrien_people_2020, lei_two-sided_2023}.

Within platform design, two tipping decisions are particularly salient: tip visibility and tip adjustability. Tip visibility refers to whether workers can see tip amounts before accepting deliveries \citep{warren_feeling_2021}, while tip adjustability determines whether customers can modify tips after service completion \citep{lei_two-sided_2023}. These features create different incentive structures and information flows that significantly influence service quality and satisfaction outcomes, yet their simultaneous consequences for both customers and workers remain poorly understood.


\begin{figure} [!ht]
    \centering
    \includegraphics[width=0.4\linewidth]{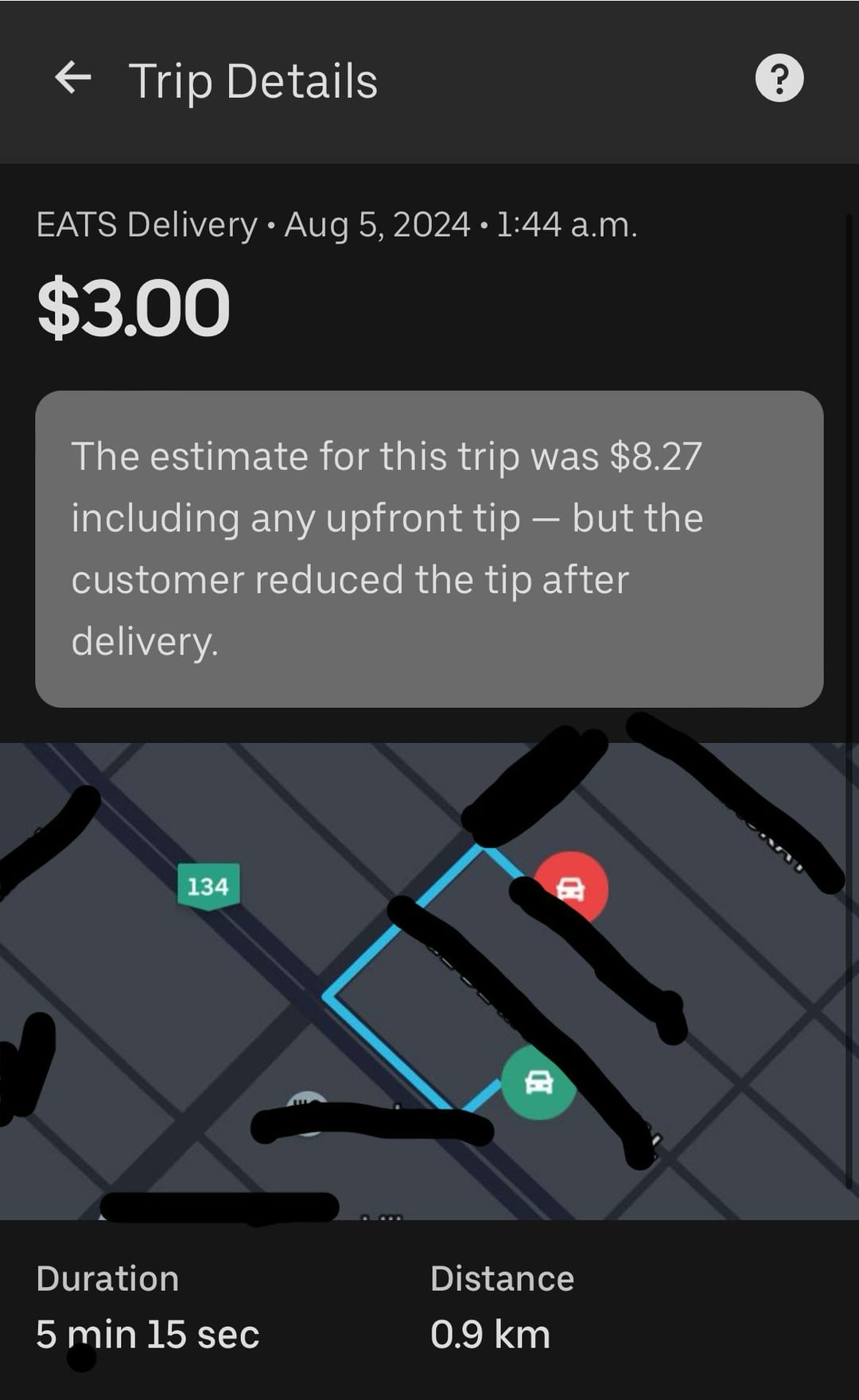}
    \caption{Example of the worker-facing app when the customer has removed a tip after food delivery completion.}
    \label{fig:tipRemoval}
\end{figure}

\subsection{Dyadic Satisfaction in Platform Operations}

Traditional logistics service quality (LSQ) research has predominantly focused on customer satisfaction \citep{mentzer_physical_1989, parasuraman_servqual_1988}. This focus has inadvertently relegated delivery workers to a commoditized position within the fulfillment system. This customer-dominant perspective, while appropriate for traditional logistics systems where delivery personnel are employees, creates tension in food delivery platforms that operate as multi-sided markets.

Food delivery platforms fundamentally differ from traditional business-to-business (B2B) contexts in that they must simultaneously create and maintain value for both customer and service worker populations \citep{rysman_economics_2009}. Consider Etsy, where the platform's success depends equally on seller satisfaction and customer satisfaction - neglecting either population would undermine the platform's viability. However, current food delivery platforms appear to have inherited the customer-dominant perspective of traditional B2B logistics strategy, despite operating in a multi-sided market. This misalignment is particularly problematic given that delivery workers are independent contractors who can choose whether and when to participate in the platform.

The LSCM literature has yet to consider food delivery customer and worker satisfaction as equivalent objectives. Addressing this theoretical gap requires elevating worker satisfaction to equivalent status with customer satisfaction through dyadic measurement. \textit{Dyadic satisfaction} simultaneously captures both customer and worker satisfaction with delivery interactions, acknowledging that platform operations require maintaining satisfaction on both sides since dissatisfied workers may leave or provide lower quality service, while dissatisfied customers may switch to competing platforms.

\subsection{Service Outcomes and Dyadic Satisfaction: Tipping Policy Effects}

Service outcomes encompass operational factors (delivery efficiency, food quality), relational factors (professional interactions), and price factors (value perception relative to compensation) \citep{rao_electronic_2011, ta_reconceptualizing_2025}. Customers evaluate these dimensions relative to their service expectations and tipping decisions, while workers evaluate them relative to expected time and effort investment and total compensation. Table~\ref{tab:dyadic_experiences} summarizes the parallel aspects of e-LSQ for customers and workers.

\begin{table}[!ht]
\centering
\caption{Parallel Service Quality Experiences in Food Delivery Dyads}
\label{tab:dyadic_experiences}
\begin{tabular}{p{3cm}p{5.5cm}p{5.5cm}}
\toprule
\textbf{e-LSQ Dimension} & \textbf{Customer Experience Aspect} & \textbf{Worker Experience Aspect} \\
\midrule
\textbf{Operational Factors} & 
Timely delivery, proper food temperature, accurate order fulfillment, appropriate packaging & 
Efficient restaurant operations, minimal wait times, orders ready upon arrival, clear delivery instructions, accessible locations \\[0.5em]

\textbf{Relational Factors} & 
Professional worker communication, courteous delivery interaction, appropriate handling of service issues & 
Respectful treatment from restaurant staff, clear customer communication about preferences, professional interactions at pickup and delivery \\[0.5em]

\textbf{Price Factors} & 
Value assessment relative to total cost including tip, perception of fair pricing for service received & 
Compensation evaluation relative to time and effort expended, consideration of distance, traffic, and delivery complexity \\
\bottomrule
\end{tabular}
\end{table}

\begin{figure} [!ht]
    \centering
    \includegraphics[width=.8\linewidth]{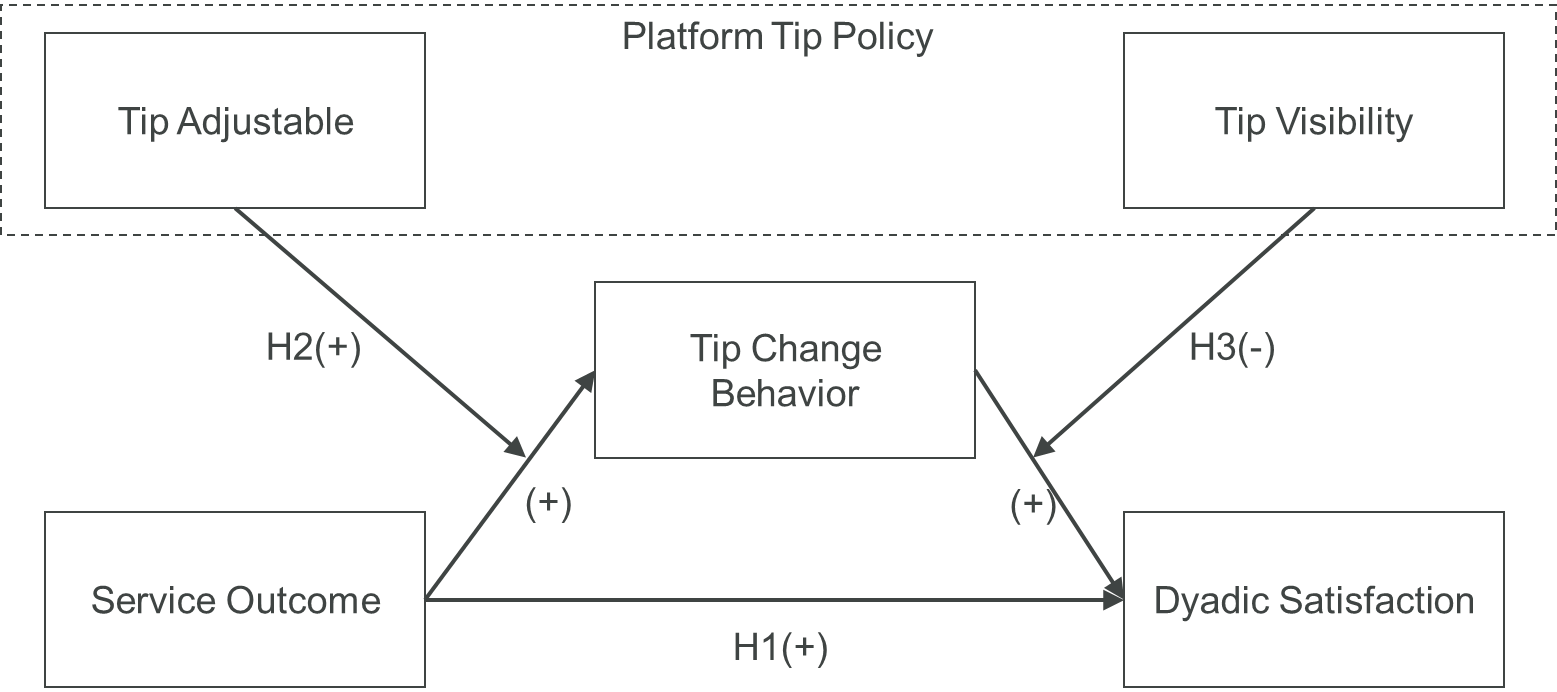}
    \caption{Hypothesized Moderated Mediation Model of Tipping Policy on Dyadic Satisfaction.}
    \label{fig:hypFramework}
\end{figure}

Research consistently shows that service quality significantly impacts satisfaction in traditional B2B logistics contexts \citep{mentzer_physical_1989, rao_electronic_2011}. In food delivery platforms, service quality simultaneously affects both customers and workers through their respective experiences of the service interaction. When operational factors align (efficient restaurant preparation, smooth navigation, timely delivery), relational factors are positive (professional, courteous interactions), and price factors are favorable (perceived value for customers, appropriate compensation for workers), both parties experience enhanced satisfaction (see Figure~\ref{fig:hypFramework}). Conversely, when service outcomes fall short in any dimension, both parties may experience reduced satisfaction. Therefore:

\begin{hyp} \label{hyp1}
Delivery service outcome is positively associated with dyadic satisfaction.    
\end{hyp}


Tip adjustability introduces a behavioral feedback mechanism in service interactions. Traditional service quality research suggests customers seek ways to reward good service and penalize poor service \citep{lynn_consumer_1993, lynn_tipping_2008}. In food delivery platforms, tip adjustability provides customers with a direct mechanism to align compensation with experienced service quality, enabling both genuine feedback (adjusting tips based on actual performance) and potential manipulation through ``tip baiting'' \citep{lei_two-sided_2023, zhang_allow_2024, yang_risks_2025}.

Without tip adjustability, customers' initial tips remain fixed regardless of service quality, nullifying any relationship between service outcomes and tip changes. When adjustability is enabled, customers can modify tips based on actual rather than anticipated service quality. Given that tip adjustability is necessary for any tip changes to occur:

\begin{hyp} \label{hyp2}
Tip adjustability amplifies the positive relationship between service outcome and tip change behavior.
\end{hyp}

The relationship between tip changes and dyadic satisfaction depends on when tip information becomes visible to workers \citep{warren_feeling_2021, warren_tipping_2025}. When tips are hidden until after delivery, the mechanism functions traditionally as post-service reward \citep{lynn_consumer_1993}, allowing customers unbiased service assessment and workers to receive unexpected recognition. However, when tips are visible before delivery, the mechanism transforms from reward to contract \citep{zhang_allow_2024}.

Visible tips create explicit expectations about delivery value for the worker. This makes tip increases less satisfying (already anticipated) and decreases more disappointing (violating implicit contracts) \citep{warren_tipping_2025}. This pre-service visibility also affects attribution: under \textit{a priori} hidden tips, customers attribute good service to worker professionalism while workers view higher tips as genuine appreciation. Under \textit{a priori} visible tips, customers may attribute good service to their upfront tip amount while workers view service quality as contractual fulfillment rather than exceeding expectations. In essence, tip visibility changes the psychological meaning of tip adjustments from genuine appreciation to contractual fulfillment, reducing their impact on satisfaction for both parties. Therefore:

\begin{hyp} \label{hyp3}
Tip visibility attenuates the relationship between tip change behavior and dyadic satisfaction.
\end{hyp}

\subsection{Methodology}
GABMs offer unique advantages for studying food delivery platform dynamics that traditional research methods cannot provide \citep{park_generative_2023}. Unlike field studies that cannot manipulate platform policies without significant operational disruption, or laboratory experiments that struggle to capture the complexity of real-world service encounters, GABMs enable controlled manipulation of platform design features while maintaining behavioral realism. The dyadic nature of food delivery interactions presents particular challenges for traditional ABM approaches that rely on predetermined decision rules. Customer and worker satisfaction are interdependent yet influenced by different information asymmetries in these interactions. GABMs address this limitation by enabling agents to process contextual nuances, emotional responses, and adaptive communication patterns through natural language reasoning. This capability makes GABMs particularly well-suited for simulating the complex psychological processes that characterize customer-worker dyadic interactions in platform economies.

\subsubsection{Conceptual Model}

\begin{figure} [!ht]
    \centering
    \includegraphics[width=1\linewidth]{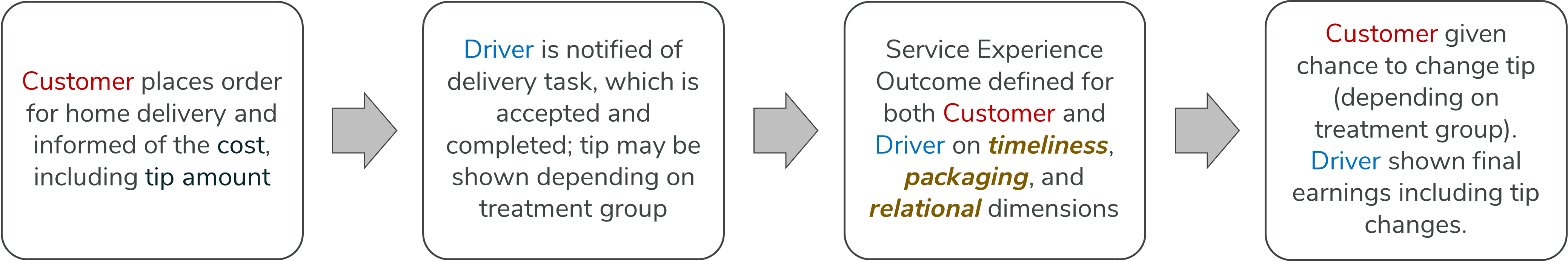}
    \caption{GABM Conceptual Model}
    \label{fig:conceptualModel}
\end{figure}

The conceptual model for this food delivery platform GABM captures the service experience sequence from both customer and worker perspectives. Figure \ref{fig:conceptualModel} illustrates the four key stages of interaction between the two platform participants. The process begins when a customer places an order and sets an initial tip amount. At this stage, the customer has full information about costs, including their chosen tip amount, but faces uncertainty about service quality.


In the second stage, a worker receives notification of the delivery task. Depending on the experimental condition (tip visibility), the worker may or may not see the customer's initial tip amount when deciding whether to accept the task. This information asymmetry represents a key platform design element that potentially influences service expectations and behavior.

The third stage encompasses the actual service experience, where both customer and worker engage in the delivery process. The service outcome is defined along three critical dimensions: timeliness (e.g., restaurant preparation speed, delivery time), packaging (e.g., food condition, order accuracy), and relational aspects (e.g., communication, courtesy). These dimensions align with established e-LSQ frameworks while accounting for both customer and worker perspectives on service quality.

The final stage represents the post-service period where customers may modify their initial tip amount depending on the experimental condition (tip adjustability). The worker then sees their final earnings, including any tip changes. This stage is crucial for understanding how the ability to adjust tips after service completion affects both customer and worker satisfaction with the platform experience.

\subsubsection{Experimental Design}

To examine how platform design features affect dyadic satisfaction, I employed a 4×2×2 factorial design that enables investigation of the effects of service outcomes, tip adjustability, and tip visibility on dyadic satisfaction. The first factor, \textit{Service Outcome}, has four levels that comprehensively capture the delivery experience. Table~\ref{tab:serviceOutcomes} lists the four levels of this variable. The descriptions were tailored to both customer and worker perspectives. Within each service outcome level, customer and worker experiences are aligned to ensure internal consistency. For instance, for food to arrive early from the customer's perspective (Exceeds), the worker must not be subjected to late pickup times at the restaurant (Fails).

\begin{table}[!ht]
\centering
\caption{Service Outcome Descriptions for Customer and Worker}
\label{tab:serviceOutcomes}
\begin{tabular}{p{2.5cm}p{6cm}p{6cm}}
\toprule
\textbf{Outcome} & \textbf{Customer Perspective} & \textbf{Worker Perspective} \\
\midrule
\textbf{Exceeds} &
Your food arrived earlier than expected. The food is in excellent condition and packaging. The worker was exceptionally polite and professional. &
After driving to the pickup location, the restaurant had the order ready, so you did not have to wait. You were able to deliver the order earlier than expected. Upon delivering at the drop-off location, the customer was exceptionally polite. \\[0.5em]

\textbf{Meets} &
Your food arrived on time. The food is in good condition and packaging. The worker was polite. &
After driving to the pickup location, the restaurant did not have the order ready, making you wait a bit longer than expected. You were able to deliver the order on time. Upon delivering at the drop-off location, the customer was polite. \\[0.5em]

\textbf{Below} &
Your food arrived slightly late. The food is somewhat cold and not well-packaged. The worker was slightly rude. &
After driving to the pickup location, the restaurant did not have the order ready, making you wait much longer than expected. You made the delivery slightly later than expected. Upon delivering at the drop-off location, the customer was slightly rude. \\[0.5em]

\textbf{Fails} &
Your food arrived very late. The food is in poor condition or incorrect packaging. The worker was very rude and unprofessional. &
After driving to the pickup location, the restaurant did not have the order ready, making you wait significantly longer than expected. You made the delivery much later than expected. Upon delivering at the drop-off location, the customer was very rude. \\
\bottomrule
\end{tabular}
\end{table}

The second factor, \textit{Tip Adjustability}, has two levels: adjustable (where customers can modify their tip up to 24 hours after delivery) and non-adjustable (where the initial tip amount cannot be changed). The option to change tips was presented to only those participants or agents that were randomly assigned to that group. 

The third factor, \textit{Tip Visibility}, has two levels: before delivery (where workers see the tip amount when deciding whether to accept a delivery) and after delivery (where workers only see tips after completing the delivery). Participants or agents acting as customers were shown the corresponding visibility information that matched their randomly assigned group.

\subsubsection{Satisfaction Data Collection}

Both LLMs and human participants completed structured surveys following the vignette presentation. Customer-role participants (or agents) in the tip-adjustable conditions were first given the opportunity to modify their initial tip amount based on the described service experience. All customers then rated their satisfaction with the delivery experience on a 7-point Likert scale (1 = extremely dissatisfied, 7 = extremely satisfied). Worker-role participants (agents) subsequently received information about the customer's final tip amount and were asked to evaluate their own satisfaction on the same 7-point scale. Both customers and workers were instructed to provide open-ended explanations for their satisfaction ratings to capture the reasoning behind their evaluations. This sequential process ensured that worker satisfaction assessments incorporated knowledge of customer behavior, reflecting the interdependent nature of dyadic service encounters.

\subsubsection{Tip Change and Dyadic Satisfaction Outcome Measures}

\textit{Tip change} captures customers' willingness to adjust tips based on service outcome experiences. This variable is calculated as the difference between customers' final tip amount and their initial tip amount $(\textit{tip change} = \textit{final tip} - \textit{initial tip})$. All LLM and human participants were shown the same total purchase price (\$30 USD) and were given the initial tip amount. For the participants assigned to the no tip adjustability group, structural zeros were imputed (\textit{tip change} = 0) to reflect the constraint that participants could not modify their gratuity regardless of service quality. This approach ensures that non-adjustable conditions contribute meaningful information about behavioral intentions under platform restrictions rather than generating missing data. The resulting tip change scores were mean-centered for analysis to create a measure where positive (negative) values indicate tip increases (decreases), and zero represents no change from the initial amount or structural inability to adjust.

The final outcome variable in this study captures \textit{dyadic satisfaction}. Food delivery experiences involve interdependent outcomes for both customers and workers. Following established dyadic analysis frameworks \citep{kenny_dyadic_2006}, I measured two types of satisfaction to comprehensively assess platform performance.

\textit{Joint satisfaction} represents the average satisfaction across the customer-worker dyad, calculated as:

\begin{equation*}
\text{Joint Satisfaction} = \frac{(\text{Customer Satisfaction} + \text{Worker Satisfaction})}{2}
\end{equation*}

This measure captures the overall success of the delivery encounter from a platform perspective. Platforms benefit when both sides of the marketplace report positive experiences \citep{wang_effects_2025}. This measure was mean-centered.

\textit{Differential satisfaction} measures the satisfaction gap within each dyad, calculated as:

\begin{equation*}
\text{Differential Satisfaction} = \text{Customer Satisfaction} - \text{Worker Satisfaction}
\end{equation*}

This metric reveals potential asymmetries in service experience outcomes and identifies conditions where one party benefits at the expense of the other. This measure was also mean-centered prior to analysis. Positive (negative) values indicate customers experienced higher (lower) satisfaction than workers. Values closer to zero indicate similar experiences while values closer to $\pm 1$ indicate dissimilarity. 

Following Law and Kelton's (1982) sample size calculations for stochastic simulation studies, I conducted pilot tests to estimate response variability across LLM agents. Using the highest standard deviation observed for satisfaction measures, I set the half-width criterion to h=1.0 satisfaction points and applied $z_{0.975}=1.96$ to calculate that each scenario combination required 30 replications to ensure adequate statistical precision. This resulted in 480 dyads needed per LLM (4 service outcomes × 2 tip adjustability levels × 2 tip visibility levels × 30 replications). This design yielded sufficient statistical power for both individual model assessment and cross-model comparisons essential for the dual-validation framework.

\subsubsection{LLM Implementation and Data Collection}

I systematically evaluated six state-of-the-art LLMs representing three major model families: OpenAI's \textit{GPT-4o} and \textit{GPT-4.1}, Anthropic's \textit{Claude Sonnet 3.5} and \textit{Claude Sonnet 4}, and Mistral AI's \textit{Large-2} and \textit{Medium-3}. This selection enabled both cross-family comparisons and within-family temporal analysis.

The LLM data collection process utilized programmatic prompt generation and API integration to ensure experimental control and reproducibility (see Figure~\ref{alg:llm_simulation_process} for pseudocode). I developed JSON-formatted vignette libraries containing standardized scenario descriptions for each experimental condition, with separate customer and worker perspectives (see Appendix~\ref{app:vignettes}). Python scripts automatically selected appropriate vignettes based on experimental conditions and constructed role-specific prompts that included scenario context, platform feature information, and structured response requirements. Detailed logs were recorded for tracing and verification purposes \citep{rand_agent-based_2011, sargent_verification_2013}.

API calls were implemented through multiple providers: Microsoft Azure for OpenAI models, AWS Bedrock for Anthropic models, and direct API integration for Mistral models. Each API call included standardized parameters (temperature=0.7, top\_p=0.95) calibrated to balance response consistency with natural behavioral variation. The system implemented robust error handling with exponential backoff retry mechanisms to ensure data completeness while accommodating API rate limits and occasional service interruptions.

Response parsing utilized regular expression (regex) matching to extract satisfaction ratings, reasoning explanations, and tip adjustment decisions from structured LLM outputs. Each response underwent immediate validation to confirm adherence to required formats and value ranges. Failed responses triggered automatic retry procedures to maintain data integrity.

\subsubsection{Human Participant Data Collection}
Human validation data was collected through Prolific to establish behavioral baselines for the dual-validation framework \citep{palan_prolificacsubject_2018}. I recruited 960 participants meeting strict quality criteria: greater than 95\% approval rating, greater than 100 previous submissions, U.S. residency, English fluency, and monthly food delivery service usage in the preceding year. This sample size satisfied the statistical power requirements established through Law and Kelton \citeyearpar{law_simulation_1982} calculations. Screenshots of scenarios presented to humans are presented and described in detail in Appendix~\ref{app:vignettes}.

Participants were randomly assigned to customer or worker roles within the same 4x2×2 experimental conditions used for LLM evaluation. The experimental protocol was written in Python and deployed via oTree \citep{chen_otreeopen-source_2016} to present identical vignette scenarios and collect structured responses including satisfaction ratings (7-point Likert scale), reasoning explanations, and tip adjustment decisions. Participants received a base compensation of \$0.25 for study completion (approximately \$15/hour based on pilot testing), with no performance-based incentives to minimize response bias and ensure authentic preference revelation.

Quality control measures included attention checks (directed responses), open-ended response completeness checks, and AI-generated response detection protocols \citep{abbey_attention_2017}. To assess the potential of participants using generative AI to complete the survey, the data collection page included a request, written in white text, to produce a haiku about food delivery within their response (in case of copying and pasting into a generative AI model) . This technique applied during pilot testing revealed AI contamination in an alternative online data collection platform, requiring a move to Prolific. Prolific's verified human participant base and quality screening procedures coupled with the stringent quality checks employed in this study ensured trustworthy human behavioral data. 

\subsection{Analysis and Results}

\subsubsection{Analysis Strategy}
All statistical analyses were conducted in Python using the following statistical packages: \textit{scipy} for descriptive statistics \citep{virtanen_scipy_2020}, \textit{pingouin} for Welch's ANOVA procedures and Games-Howell pairwise comparisons \citep{vallat_pingouin_2018}, and \textit{semopy} for SEM analysis \citep{georgy_semopy_2020}. I replaced means-difference testing as in conventional null hypothesis significance testing (NHST) with equivalence testing. However, Welch’s ANOVA with Games–Howell comparisons are included to illustrate the limitation of NHST for assessing differences between LLM and human behavioral responses.

First, equivalence testing on each of the three outcome variables was conducted using a series of Two One-Sided Tests (TOST), which examine whether each LLM’s mean response falls within a pre-specified equivalence margin ($\pm 0.2 SD$) around the human mean \citep{lakens_equivalence_2017, lakens_equivalence_2018, rainey_arguing_2014}. Unlike NHST where non-significant results only imply a failure to detect an effect, TOST can affirm practical equivalence by rejecting the hypothesis that the difference between AI and humans exceeds the equivalence margin. This makes TOST appropriate when the goal is to show that an AI behaves similarly to humans rather than merely to show that it is not significantly different, as in NHST.

Second, beyond surface-level equivalence, the complete dyadic decision pathway is modeled using multi-group SEM with bias-corrected bootstrap confidence intervals based on 5000 resamples \citep{shrout_mediation_2002, efron_introduction_1994, hayes_introduction_2017}. SEM is required because the theorized model includes double moderated mediation, which should not be separated with standalone regressions. The multi-group (six LLM groups and one human group) setup allows for testing whether the same paths and coefficients hold for each of the seven groups. Bootstrapping, which supplies distribution-free standard errors and confidence intervals, is needed in case the theorized indirect effects do not follow a normal sampling distribution. Together, the SEM with bootstrapping moves the analysis from ``does AI match the human outcome?'' to ``does AI follow the same psychological process that produces the human outcome?''

\subsubsection{Sample Descriptive Statistics}

The final human sample consisted of 477 dyads (957 individual participants) after applying quality screening procedures including attention check validation and AI-generated response detection. Three dyads were removed due to three human participants failing quality checks, making that participant's randomly assigned dyad unusable. The AI sample comprised 2,880 dyads (5,760 individual responses) distributed evenly across six LLMs with 480 dyads per model. Each AI dyad consisted of two separate LLM instances (one customer agent, one worker agent) using identical model parameters to ensure experimental consistency. Table~\ref{tab:descriptive_stats} reports the descriptive statistics of the outcome variables.

\begin{table}[htbp]
\centering
\caption{Descriptive Statistics: Tip Change, Joint Satisfaction, and Differential Satisfaction}
\label{tab:descriptive_stats}
\begin{threeparttable}
\begin{tabular}{l*{6}{S[table-format=3.2]}}
\toprule
\multirow{2}{*}{Model} & \multicolumn{2}{c}{Tip Change} & \multicolumn{2}{c}{Joint Satisfaction} & \multicolumn{2}{c}{Differential Satisfaction} \\
\cmidrule(lr){2-3} \cmidrule(lr){4-5} \cmidrule(lr){6-7}
& {M} & {SD} & {M} & {SD} & {M} & {SD} \\
\midrule
Human            & -0.08 & 3.61 & 0.07 & 2.08 & -0.42 & 2.11 \\
GPT-4o           & 0.15  & 3.42 & 0.07 & 2.27 & -0.37 & 1.38 \\
GPT-4.1          & -0.87 & 5.32 & 0.20 & 2.14 & -0.02 & 1.49 \\
Sonnet 3.5       & 0.11  & 3.34 & -0.15 & 2.12 & -0.29 & 1.37 \\
Sonnet 4         & 0.30  & 4.16 & -0.02 & 2.03 & 0.29 & 1.26 \\
Mistral Large 2  & 0.07  & 3.98 & -0.05 & 2.15 & 0.34 & 1.23 \\
Mistral Medium 3 & 0.40  & 3.74 & -0.12 & 2.23 & 0.46 & 1.17 \\
\bottomrule
\end{tabular}
\begin{tablenotes}
\small
\item \textit{Note:} All variables are mean-centered. Sample sizes: $N_{AI}=480$ dyads for each LLM and $N_{Human}=477$ dyads.
\end{tablenotes}
\end{threeparttable}
\end{table}

Several notable patterns emerged across the outcome measures. For \textit{tip change} behavior, most LLMs exhibited small positive mean adjustments or tip increases, with \textit{Mistral Medium 3} showing the largest average increase (M = 0.40) and \textit{GPT-4.1} demonstrating the only substantial negative adjustment (M = -0.87). \textit{GPT-4.1} also displayed the highest variability in tipping behavior (SD = 5.32), while \textit{Sonnet 3.5} showed the most consistent responses (SD = 3.34). Human participants exhibited slightly negative average tip changes (M = -0.08) with moderate variability (SD = 3.61). 

\begin{figure}[!ht]
    \centering
    \includegraphics[width=1\linewidth]{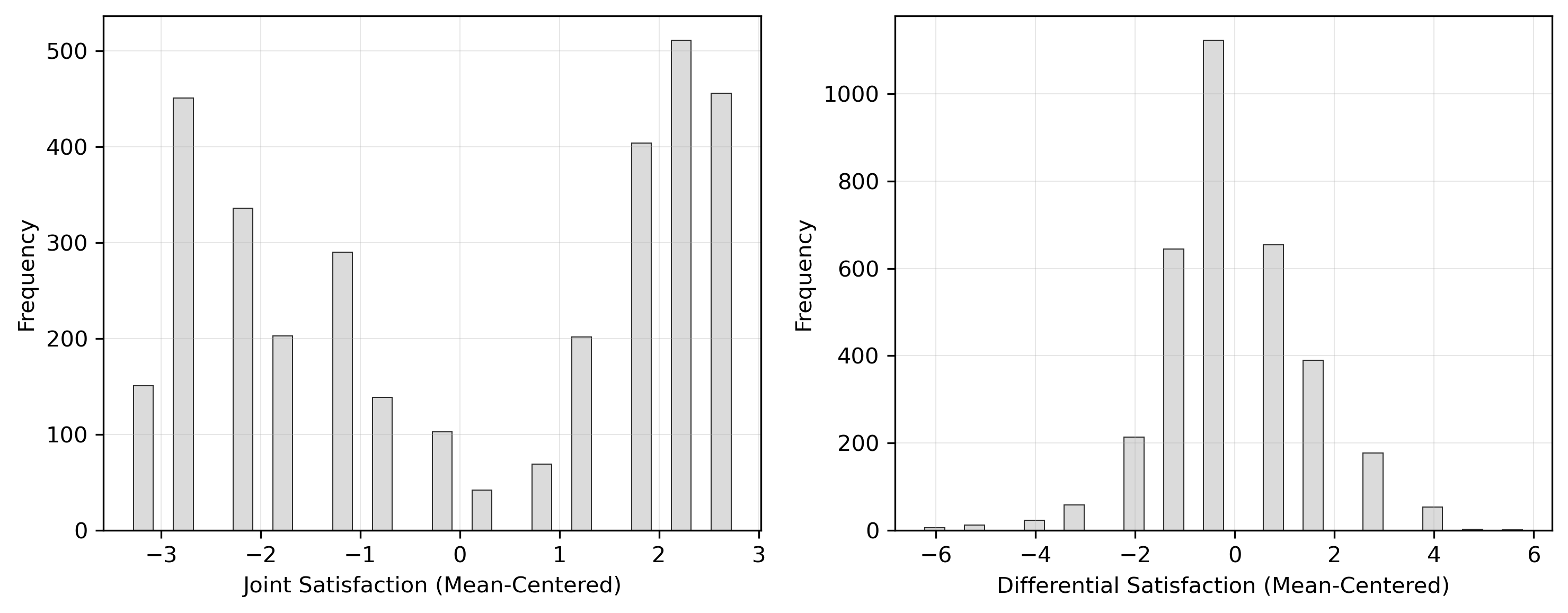}
    \caption{Overall Joint and Differential Dyadic Satisfaction Distributions}
    \label{fig:dyadicSatHist}
\end{figure}

\textit{Joint satisfaction} means clustered near zero across all groups (see Figure~\ref{fig:dyadicSatHist}), reflecting the mean-centering procedure, with relatively similar standard deviations indicating comparable overall variability. \textit{Differential satisfaction} patterns revealed more pronounced between-group differences: \textit{GPT-4o} and human participants showed negative means ($M=-0.37$ and $M=-0.42$, respectively), suggesting workers experienced higher satisfaction than customers in these groups. Conversely, both Mistral models demonstrated positive \textit{differential satisfaction} (0.34 and 0.46), indicating customers reported higher satisfaction than workers. Notably, human participants exhibited substantially greater variability in \textit{differential satisfaction} ($SD = 2.11$) compared to all AI models, suggesting more heterogeneous satisfaction experiences within human dyads.


\subsubsection{Surface-Level Behavioral Equivalence}

Next, I conducted a means-difference analysis for a traditional NHST effort. However, Levene's test revealed significant violations of homogeneity of variances for all variables: \textit{tip change} ($W = 10.18, p < 0.001$), \textit{joint satisfaction} ($W = 11.96, p < 0.001$), and \textit{differential satisfaction} ($W = 18.59, p < 0.001$). Therefore, I applied a Welch's ANOVA procedure, which revealed significant differences across participant groups (all AI models and Humans) for \textit{tip change} behavior ($F (6, 1428.41)=3.438, p=0.002$, $\eta_p^2=0.009$) and \textit{differential satisfaction} ($F(6,1485.13) = 33.064, p < 0.001$, $\eta_p^2=0.051$), while \textit{joint satisfaction} showed no significant differences ($F(6,1488.70) = 1.526, p = 0.166$, $\eta_p^2=0.003$). Games-Howell pairwise comparisons on each AI model and human pairings are reported in Appendix \ref{app:GHcomparisons}. These results indicated that models varied systematically in the \textit{tip change} and \textit{differential satisfaction} behavioral outputs, but no significant difference was found for \textit{joint satisfaction}. However, failure to reject the null hypothesis that there is no difference in participant groups' \textit{joint satisfaction} is not necessarily evidence that participant groups are similar, so equivalence testing is required.

Therefore, I also conducted TOST equivalence testing with $\pm 0.2$ standard deviation bounds to assess similarity between each LLM and human participants across the three measures (see Table~\ref{tab:tost_equivalence}). \textit{GPT-4o} was the only AI model to achieve statistical equivalence on all three measures (\textit{Tip change}: $p = 0.0185$, \textit{Joint satisfaction}: $p = 0.0011$, \textit{Differential satisfaction}: $p = 0.0041$). \textit{Mistral Large 2} and \textit{Sonnet 3.5} each achieved equivalence on two of three measures. The remaining models (\textit{GPT-4.1}, \textit{Sonnet 4}, and \textit{Mistral Medium 3}) achieved equivalence on only one measure, indicating poor surface-level behavioral fidelity. These results establish a surface-level performance hierarchy: \textit{GPT-4o}, followed by \textit{Mistral Large 2} and \textit{Sonnet 3.5} with moderate equivalence, while newer model generations consistently failed to outperform their predecessors.

\begin{sidewaystable}
\centering
\caption{TOST Equivalence Tests vs. Human Baseline (±0.2 SD)}
\label{tab:tost_equivalence}
\begin{tabular}{ll*{3}{S[table-format=-1.4]}l}
\toprule
\textbf{Comparison} & \textbf{DV} & {\textbf{Lower Bound}} & {\textbf{Upper Bound}} & {\textbf{p-value}} & \textbf{Is Equivalent} \\
\midrule
GPT-4o vs. Human & Tip change & -0.7032 & 0.7032 & 0.0185 & True \\
GPT-4o vs. Human & Joint satisfaction & -0.4357 & 0.4357 & 0.0011 & True \\
GPT-4o vs. Human & Differential satisfaction & -0.3567 & 0.3567 & 0.0041 & True \\
GPT-4.1 vs. Human & Tip change & -0.9092 & 0.9092 & 0.3430 & False \\
GPT-4.1 vs. Human & Joint satisfaction & -0.4220 & 0.4220 & 0.0140 & True \\
GPT-4.1 vs. Human & Differential satisfaction & -0.3657 & 0.3657 & 0.6113 & False \\
Sonnet 3.5 vs. Human & Tip change & -0.6953 & 0.6953 & 0.0124 & True \\
Sonnet 3.5 vs. Human & Joint satisfaction & -0.4202 & 0.4202 & 0.0756 & False \\
Sonnet 3.5 vs. Human & Differential satisfaction & -0.3558 & 0.3558 & 0.0273 & True \\
Sonnet 4 vs. Human & Tip change & -0.7758 & 0.7758 & 0.0611 & False \\
Sonnet 4 vs. Human & Joint satisfaction & -0.4110 & 0.4110 & 0.0090 & True \\
Sonnet 4 vs. Human & Differential satisfaction & -0.3474 & 0.3474 & 0.9994 & False \\
Mistral Large 2 vs. Human & Tip change & -0.7596 & 0.7596 & 0.0066 & True \\
Mistral Large 2 vs. Human & Joint satisfaction & -0.4225 & 0.4225 & 0.0137 & True \\
Mistral Large 2 vs. Human & Differential satisfaction & -0.3456 & 0.3456 & 0.9999 & False \\
Mistral Medium 3 vs. Human & Tip change & -0.7340 & 0.7340 & 0.1499 & False \\
Mistral Medium 3 vs. Human & Joint satisfaction & -0.4308 & 0.4308 & 0.0446 & True \\
Mistral Medium 3 vs. Human & Differential satisfaction & -0.3409 & 0.3409 & 1.0000 & False \\
\bottomrule
\end{tabular}
\end{sidewaystable}

\subsubsection{Process-Level Validation}

The multi-group approach allows us to estimate the SEM for the human group and each LLM. Using the SEM for human participants as the baseline (see Table~\ref{tab:sem_humans} in Appendix~\ref{app:SEMresults}), significant direct effects across multiple paths are found. \textit{Service outcome} strongly predicted \textit{tip change} behavior ($\beta = 1.169$, $SE = 0.160$, $p < 0.001$), indicating that better service experiences led to larger tip adjustments. Thus, \textbf{H1 was supported}. In the human participants model, the hypothesized moderating effect of \textit{tip adjustability} was not supported ($\beta = 0.104$, $SE = 0.163$, $p = 0.526$). Thus, \textbf{H2 was not supported}. 

\textit{Tip change} behavior significantly influenced both satisfaction outcomes. Larger tip changes were associated with higher \textit{joint satisfaction} ($\beta = 0.051$, $SE = 0.013$, $p < 0.001$) and greater \textit{differential satisfaction} ($\beta = 0.063$, $SE = 0.025$, $p = 0.012$). The moderation by \textit{tip visibility} showed divergent effects: while the interaction was non-significant for \textit{joint satisfaction} ($\beta = 0.017$, $SE = 0.018$, $p = 0.345$), it significantly affected \textit{differential satisfaction} ($\beta = -0.175$, $SE = 0.033$, $p < 0.001$), suggesting that \textit{tip visibility} reduces the positive impact of \textit{tip changes} on customer-worker satisfaction gaps. Thus, \textbf{H3 was partially supported}.

\textit{Service outcome} also exhibited strong direct effects on both satisfaction measures: \textit{joint satisfaction} ($\beta = 1.557$, $SE = 0.045$, $p < 0.001$) and \textit{differential satisfaction} ($\beta = 0.936$, $SE = 0.085$, $p < 0.001$). However, the indirect effects were not supported in the human data (see results in Table~\ref{tab:bootstrap_all}). Bootstrap analysis with 5,000 samples revealed non-significant indirect effects on both \textit{joint satisfaction} ($\beta = 0.002$, 95\% CI [-0.004, 0.013], $p = 0.606$) and \textit{differential satisfaction} ($\beta = -0.018$, 95\% CI [-0.056, 0.016], $p = 0.287$). These findings indicate that while \textit{tip adjustability} moderates the relationship between \textit{service outcomes} and \textit{tip change} behavior, this moderation does not translate into meaningful indirect effects on satisfaction outcomes in the human participants group.

In stark contrast, all AI models exhibited deviations from the human decision-making pattern when evaluated across the complete SEM structure (Tables~\ref{tab:sem_GPT4o}-\ref{tab:sem_mistral_medium}). Comprehensive analysis of all eight direct pathways and two indirect effects revealed that four models—\textit{GPT-4.1}, \textit{Sonnet 3.5}, \textit{Sonnet 4}, and \textit{Mistral Medium 3}—achieved the highest process-level fidelity with 8/10 pathway matches to human decision-making patterns (see Table~\ref{tab:model_rankings}). Surprisingly, \textit{Mistral Large 2}, despite correctly replicating the human pattern of non-significant indirect effects, ranked lower (6/10 matches) due to substantial deviations in direct pathway significance patterns. \textit{GPT-4o} demonstrated the worst process-level validation (5/10 matches), exhibiting both artificial indirect effects and multiple direct pathway deviations not present in human participants. These findings highlight that surface-level behavioral equivalence bears little relationship to underlying decision-process fidelity, with the best surface-level performer (\textit{GPT-4o}) ranking last in process validation, while several moderate surface-level performers achieved the highest process-level fidelity.

\begin{table}[!ht]
\centering
\caption{AI Model Performance Rankings Across Validation Criteria}
\label{tab:model_rankings}
\begin{tabular}{c>{\raggedright}p{3cm}>{\raggedright}p{3cm}>{\raggedright}p{3cm}>{\raggedright\arraybackslash}p{3cm}}
\toprule
\multirow{2}{*}{\textbf{Rank}} & \multicolumn{2}{c}{\textbf{Surface-Level Equivalence}} & \multicolumn{2}{c}{\textbf{Process-Level Validation}} \\
\cmidrule(lr){2-3} \cmidrule(lr){4-5}
& \textbf{AI Model} & \textbf{Measures Achieved} & \textbf{AI Model} & \textbf{Model Fidelity} \\
\midrule
1 & GPT-4o & All 3 measures & GPT-4.1 & 8/10 pathway matches \\
2 & Mistral Large 2 & 2 measures (tip, joint) & Sonnet 3.5 & 8/10 pathway matches \\
2 & Sonnet 3.5 & 2 measures (tip, diff) & Sonnet 4 & 8/10 pathway matches \\
4 & GPT-4.1 & 1 measure (joint only) & Mistral Medium 3 & 8/10 pathway matches \\
4 & Sonnet 4 & 1 measure (joint only) & Mistral Large 2 & 6/10 pathway matches \\
4 & Mistral Medium 3 & 1 measure (joint only) & GPT-4o & 5/10 pathway matches \\
\bottomrule
\end{tabular}
\begin{tablenotes}
\small
\item \textit{Note:} Surface-level equivalence based on TOST equivalence tests. Process-level validation based on SEM pathway fidelity to human decision-making patterns across 8 direct paths and 2 indirect effects. Tip = tip change; Joint = joint satisfaction; Diff = differential satisfaction.
\end{tablenotes}
\end{table}

\section{Discussion}

This study introduced Generative Agent-Based Modeling (GABM) as a novel approach for studying emergent outcomes of complex LSCM behaviors. It developed a dual-validation framework to assess the degree to which LLM-powered agents authentically replicate human decision-making outcomes and processes. Drawing from established simulation validation principles in OR and computer simulation disciplines, I have proposed both surface-level behavioral equivalence testing and process-level pathway validation as options for evaluating AI fidelity to human behavior. A food delivery tipping experiment demonstrated both the GABM methodology and the critical importance of validation. This case study reveals that surface-level performance does not guarantee alignment with underlying decision-processes.

\subsection{Theoretical Implications}

\subsubsection{Dual-Validation Framework for GABM Research}

This study's primary theoretical contribution is the development of a dual-validation framework that addresses a critical gap in GABM methodology. Traditional simulation validation approaches focus primarily on output validation that compares model results to real-world data, but these fail to address whether AI agents employ authentic decision-making processes. Recent GABM studies have called for frameworks to accomplish such a task \citep[e.g.][]{lu_llms_2024, larooij_large_2025}. The dual-validation framework contributes to GABM and simulation methodology by recognizing that LLM-powered agents require validation at two distinct levels: surface-level behavioral equivalence and process-level decision authenticity.

The framework's theoretical significance lies in establishing that these validation levels serve different research purposes and may yield conflicting results. Surface-level validation suffices for applications requiring aggregate behavioral prediction, pattern recognition, or routine decision support. Process-level validation becomes essential when understanding causal mechanisms matters—including theory development, policy analysis, and strategic decision-making. This distinction fundamentally changes how researchers should approach GABM validation, moving beyond the traditional assumption that behavioral equivalence implies process authenticity.

The finding that surface-level and process-level performance may not correspond to each other challenges existing simulation validation paradigms. The disconnect between these validation levels establishes that traditional validation approaches are insufficient for GABM applications. This contribution has broader implications for AI-powered simulation across disciplines, suggesting that any field employing LLMs for behavioral modeling must validate both what models produce and how they produce it.

\subsubsection{Dyadic Satisfaction and Dyad-Centric Platform Design}

The GABM case study contributes empirically by extending consumer-centric platform design into dyad-centric design. The research demonstrates that platform satisfaction operates as a dyadic phenomenon. Traditional platform research treats workers as service providers whose satisfaction matters primarily for retention and quality maintenance \citep{zhou_platform_2022}. This study establishes worker satisfaction as equally fundamental to platform operations, with both customer and worker experiences simultaneously shaped by the same operational factors.

The parallel experiences framework reveals that service quality improvements create value for both dyad members through different but complementary mechanisms. Efficient restaurant operations reduce customer wait times while minimizing worker idle time; accurate orders enhance customer satisfaction while reducing worker stress from handling complaints. This finding theoretically advances platform design by showing that dyad-centric approaches can simultaneously optimize outcomes for both parties rather than requiring trade-offs between customer and worker satisfaction.

The tip visibility findings contribute to behavioral LSCM theory by demonstrating how information timing affects satisfaction alignment within dyads. The reduced impact of tip changes on satisfaction gaps when tips are visible after (rather than before) service suggests that obligation framing significantly influences worker satisfaction dynamics. This extends prospect theory and fairness perception research into platform-mediated service contexts, showing how information presentation timing can be strategically used to promote dyadic satisfaction alignment \citep{zhang_allow_2024}.

\subsubsection{Research Agenda for GABMs in LSCM}

GABMs represent a nascent research paradigm with significant potential across multiple LSCM domains. In supply network resilience research, GABMs could simulate how individual firm decision-making under uncertainty propagates through complex networks, moving beyond aggregate shock models to examine heterogeneous responses and adaptive behaviors. Humanitarian logistics research could employ GABMs to model beneficiary behavior, local community dynamics, and cultural factors that influence aid distribution effectiveness—capabilities that current ABMs cannot adequately capture.

Last-mile delivery research represents particularly promising territory, where GABMs could simulate deeper interactions between customers, drivers, and platform algorithms across diverse market environments. GABMs could incorporate customer availability patterns, driver preferences, and real-time decision-making that emerges from complex behavioral interactions, enabling researchers to assess how well established vehicle routing methods perform with enhanced empirical realism. Similarly, sustainable supply chain research could use GABMs to model how individual consumer environmental preferences aggregate into market-level demand shifts, enabling more realistic assessment of sustainability intervention effectiveness.

While this study demonstrated GABM validation using TOST equivalence testing for surface-level validation and SEM for process-level validation, the framework is method-agnostic. Researchers should consider various statistical and analytical techniques for establishing both validation levels depending on their specific research context and objectives. Table~\ref{tab:gabm_validation_methods} provides examples of alternative validation approaches across different LSCM research domains.

\begin{table}[!htbp]
\centering
\caption{GABM Validation Methods for LSCM Applications}
\label{tab:gabm_validation_methods}
\begin{threeparttable}
\footnotesize
\begin{tabularx}{\textwidth}{>{\raggedright\arraybackslash}p{3.5cm}>{\raggedright\arraybackslash}X>{\raggedright\arraybackslash}X}
\toprule
\textbf{LSCM Research Domain} & \textbf{Surface-Level Validation Methods} & \textbf{Process-Level Validation Methods} \\
\midrule
\textbf{Supply Network Resilience} & 
\makecell[l]{$\bullet$ TOST equivalence testing of \\ disruption recovery times \\ $\bullet$ Kolmogorov-Smirnov tests \\ comparing firm survival rates \\ $\bullet$ Chi-square tests for supplier \\ switching patterns} & 
\makecell[l]{$\bullet$ SEM modeling of risk perception \\ $\rightarrow$ adaptation behavior pathways \\ $\bullet$ Mediation analysis of stress $\rightarrow$ \\ decision $\rightarrow$ outcome chains \\ $\bullet$ Network analysis of trust \\ propagation mechanisms} \\
\midrule
\textbf{Humanitarian Logistics} & 
\makecell[l]{$\bullet$ Mann-Whitney U tests \\ comparing aid distribution efficiency \\ $\bullet$ Effect size measures for \\ beneficiary satisfaction scores \\ $\bullet$ Graphical comparison of \\ resource allocation patterns} & 
\makecell[l]{$\bullet$ Path analysis of cultural factors \\ $\rightarrow$ acceptance $\rightarrow$ distribution \\ effectiveness \\ $\bullet$ Regression-based mediation of \\ community trust pathways \\ $\bullet$ Hierarchical modeling of nested \\ cultural/geographic effects} \\
\midrule
\textbf{Last-Mile Delivery} & 
\makecell[l]{$\bullet$ TOST testing of delivery \\ time distributions \\ $\bullet$ Bayesian equivalence testing \\ for customer satisfaction \\ $\bullet$ Robust statistical methods \\ for driver behavior patterns} & 
\makecell[l]{$\bullet$ SEM of platform algorithm $\rightarrow$ \\ driver decision $\rightarrow$ customer experience \\ $\bullet$ Bootstrap confidence intervals \\ for indirect satisfaction effects \\ $\bullet$ Causal inference methods for \\ pricing impact pathways} \\
\midrule
\textbf{Sustainable Supply Chains} & 
\makecell[l]{$\bullet$ $t$-tests comparing environmental \\ preference aggregation \\ $\bullet$ Distribution comparisons of \\ purchasing decisions \\ $\bullet$ Simple correlation analysis for \\ sustainability adoption rates} & 
\makecell[l]{$\bullet$ Structural equation modeling of \\ values $\rightarrow$ preferences $\rightarrow$ behavior \\ $\bullet$ Mediation analysis of information \\ $\rightarrow$ attitude $\rightarrow$ purchase intention \\ $\bullet$ Network analysis of peer \\ influence on green choices} \\
\midrule
\textbf{Buyer-Supplier Negotiations} & 
\makecell[l]{$\bullet$ Equivalence testing of \\ negotiation outcomes \\ $\bullet$ Effect size measures for \\ relationship satisfaction \\ $\bullet$ Non-parametric tests for \\ concession patterns} & 
\makecell[l]{$\bullet$ Process validation of trust $\rightarrow$ \\ strategy $\rightarrow$ outcome pathways \\ $\bullet$ SEM of power dynamics and \\ relationship quality \\ $\bullet$ Path analysis of information \\ sharing mechanisms} \\
\midrule
\textbf{Platform Ecosystem Dynamics} & 
\makecell[l]{$\bullet$ TOST testing of worker \\ retention rates \\ $\bullet$ Distributional comparisons \\ of earnings patterns \\ $\bullet$ Statistical tests for geographic \\ coverage equivalence} & 
\makecell[l]{$\bullet$ SEM modeling of service quality \\ $\rightarrow$ satisfaction $\rightarrow$ loyalty chains \\ $\bullet$ Mediation analysis of platform \\ features $\rightarrow$ behavior $\rightarrow$ outcomes \\ $\bullet$ Network analysis of worker-customer \\ interaction patterns} \\
\bottomrule
\end{tabularx}
\begin{tablenotes}
\footnotesize
\item \textit{Note:} Surface-level methods focus on behavioral outcome equivalence using statistical comparisons. Process-level methods examine underlying decision pathways and causal mechanisms. Method selection should align with research objectives and operational requirements.
\end{tablenotes}
\end{threeparttable}
\end{table}

Beyond standalone applications, GABMs offer significant potential for hybrid simulation models where agents can dynamically navigate simulated environments using natural language reasoning. For example, worker agents experiencing poor service conditions could autonomously decide to leave a platform or relocate to different market regions based on their accumulated experiences. This capability enables simulation of platform ecosystems where agent movement and decision-making create realistic market dynamics, worker turnover patterns, and geographic service coverage variations that emerge from individual behavioral choices rather than predetermined rules.

The methodology's greatest theoretical potential lies in studying emergent phenomena—system-level outcomes that arise from complex behavioral interactions but cannot be predicted from individual behaviors alone. GABMs enable researchers to bridge the micro-macro divide that has long challenged LSCM research, allowing investigation of how individual psychological processes scale up to create supply chain disruptions, market dynamics, and competitive advantages. This capability positions GABMs as a transformative methodology for advancing behavioral LSCM theory development.

\subsection{Managerial Implications}

The dual-validation framework provides practical guidance for selecting and deploying LLMs in operational contexts as well. Organizations can now make evidence-based decisions about which validation level their specific applications require. For agentic applications involving strategic decision-making, stakeholder relationship management, or novel scenario planning, models demonstrating superior process-level validation may be essential despite moderate surface-level performance. However, for routine operational tasks such as demand forecasting, pattern recognition, or basic customer service interactions, surface-level equivalence may be sufficient, allowing organizations to deploy more readily available or cost-effective models.

This framework also enables responsible AI adoption by providing systematic criteria for model selection rather than relying on general performance benchmarks that may not reflect operational requirements. Organizations can avoid over-engineering their AI deployments by matching validation requirements to specific use cases, potentially reducing implementation costs while ensuring appropriate behavioral fidelity for their operational context.

\subsection{Limitations and Future Research}

This simulation-forward research effort, grounded in operations research and computational simulation methodologies, represents an initial exploration of GABM validation. Future studies should extend beyond vignette-based experimental designs to incorporate more complex behavioral paradigms that capture richer decision-making contexts and longer-term behavioral patterns.

Most critically, this study examined AI responses to static vignettes rather than dynamic operational simulations. Future research should investigate how generative agents perform in operational environments such as last-mile delivery simulations where both worker and customer autonomy significantly influence system outcomes. Such studies would examine whether the validation framework holds when agents must navigate complex, evolving scenarios with multiple stakeholders and real-time decision-making requirements.

An important methodological limitation involves the alignment of customer and worker experiences within each dyad for internal consistency purposes. In reality, service encounters often involve asymmetric experiences where customers and workers may have markedly different perceptions of the same interaction due to role-specific concerns, information asymmetries, or contextual factors. Future research should explore scenarios where customer and worker experiences systematically diverge to better understand how GABMs handle conflicting perspectives and competing interests within dyadic relationships.

Additionally, research should explore cross-domain validation to determine whether LLM performance patterns observed in food delivery contexts generalize to other LSCM dyadic relationships, including buyer-supplier negotiations, shipper-carrier interactions, and humanitarian aid distribution scenarios.

\singlespacing
\setlength\bibsep{0pt}
\bibliographystyle{apalike}
\bibliography{references}

\titleformat{\section}[block]{\normalfont\Large\bfseries}{\appendixname~\Alph{section} - }{0em}{}
\appendix
\pagenumbering{roman}  
\setcounter{page}{1}   

\setcounter{figure}{0}
\setcounter{table}{0}
\renewcommand{\thefigure}{\thesection.\arabic{figure}}
\renewcommand{\thetable}{\thesection.\arabic{table}}

\counterwithin{figure}{section}
\counterwithin{table}{section}

\doublespacing

\section{Vignette Examples and LLM Implementation Technical Details} \label{app:vignettes}

\subsection{Human Participant Survey Interface and AI Prompt Generation}

The experiment maintained strict equivalence between human and AI conditions by using identical standardized vignettes stored in JSON format files. Each vignette contained detailed food delivery scenarios that varied systematically across the experimental manipulations: tip adjustability (TRUE/FALSE), tip visibility (before/after), and service outcomes (fails/below/meets/exceeds expectations).

Human participants accessed these scenarios through a web-based survey platform that presented the vignette text within a structured interface. The same scenario descriptions were directly extracted from the JSON files and formatted as natural language prompts for AI models. This approach ensured that both human participants and AI models received identical situational information, differing only in presentation medium rather than content.

\begin{figure} [!ht]
    \centering
    \includegraphics[width=1\linewidth]{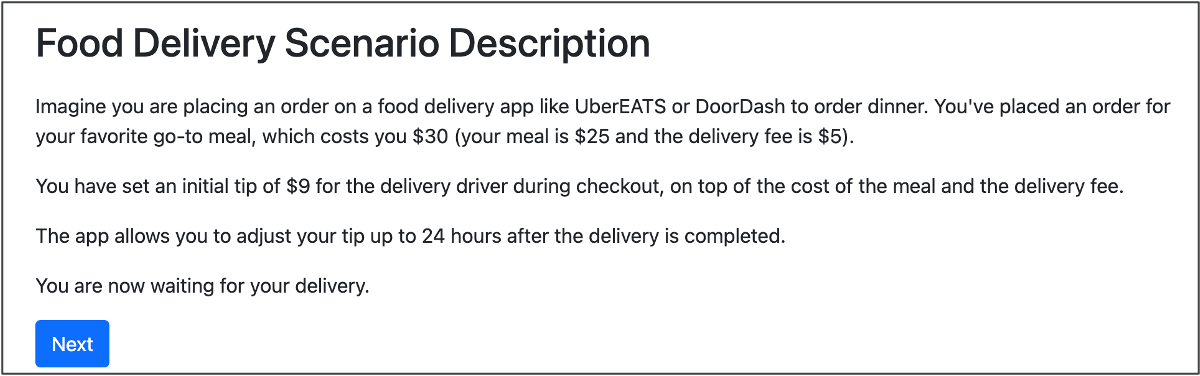}
    \caption{Customer scenario interface showing standardized vignette presentation with experimental condition details. The scenario text was extracted directly from JSON files and converted to structured prompts for AI models to ensure identical presentation across human and AI participants.}
    \label{fig:custScenario}
\end{figure}

\begin{figure} [!ht]
    \centering
    \includegraphics[width=1\linewidth]{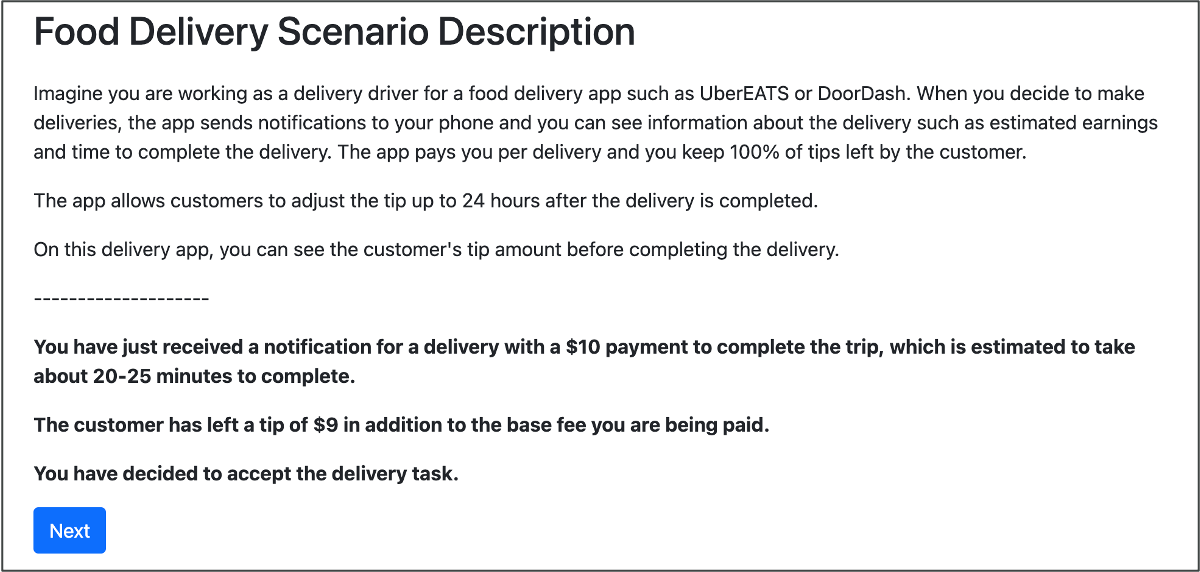}
    \caption{Worker scenario interface displaying the same delivery encounter from the driver perspective. This role-specific vignette text formed the basis for AI prompts to ensure identical presentation as humans while providing contextually appropriate information for each dyad member.}
    \label{fig:workerScenario}
\end{figure}

The vignette structure included role-specific contextual information, with customers receiving details about their initial tip amount and tip modification options (Figure~\ref{fig:custScenario}), while workers received information about tip visibility timing and delivery completion notifications (Figure~\ref{fig:workerScenario}). For example, a customer vignette might describe placing a \$9.00 initial tip during checkout, experiencing a service failure, and then being informed about their ability to modify the tip post-delivery. The corresponding driver vignette described the same delivery encounter from the driver's perspective, including when they became aware of the tip amount relative to accepting or completing the delivery task. Participants were then shown a new page describing the service outcome (see Table~\ref{tab:serviceOutcomes}).

\begin{figure} [!ht]
    \centering
    \includegraphics[width=0.6\linewidth]{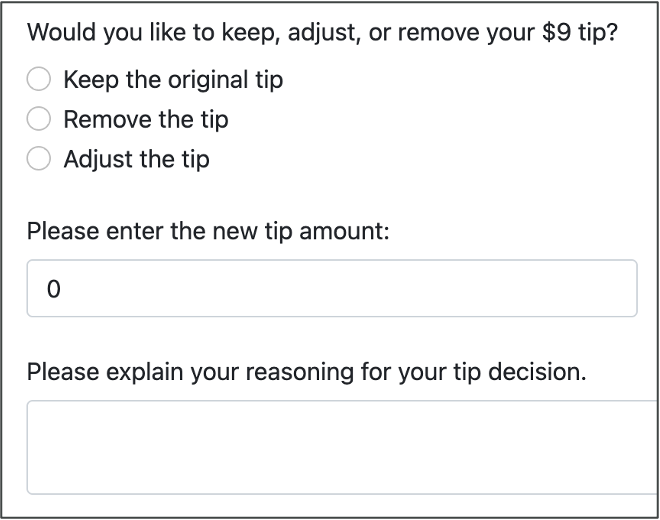}
    \caption{Customer tip modification interface shown only in tip-adjustable experimental conditions. The decision options (keep/remove/adjust) and amount specification field were translated into structured prompt instructions for AI models using identical response format requirements.}
    \label{fig:tipChange}
\end{figure}

\begin{figure} [!ht]
    \centering
    \includegraphics[width=0.6\linewidth]{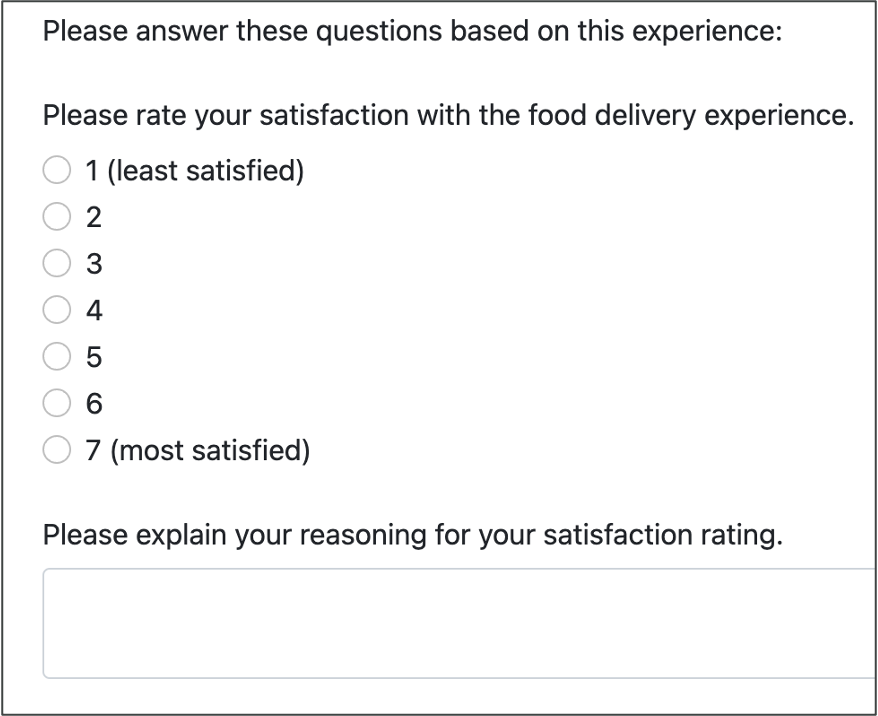}
    \caption{Satisfaction evaluation interface used by both customer and worker participants. The 7-point Likert scale and open-ended reasoning request were converted to structured formatting instructions in AI prompts to enable direct comparison of satisfaction responses across human and AI participants.}
    \label{fig:satisfactionSurvey}
\end{figure}

Response collection followed identical structured formats across both human and AI participants. In tip-adjustable conditions, customers made decisions to keep, remove, or adjust their initial tips, with specific amount modifications when applicable (Figure~\ref{fig:tipChange}). All participants provided satisfaction ratings on a 7-point Likert scale (1=very dissatisfied, 7=very satisfied) and explanatory reasoning for their ratings (Figure~\ref{fig:satisfactionSurvey}). The AI prompts included explicit formatting instructions mirroring the human survey structure: \texttt{SATISFACTION: [rating]} and \texttt{REASONING: [explanation]} for all participants, with additional \texttt{TIP DECISION: [keep/remove/adjust]} and \texttt{NEW TIP AMOUNT: [amount]} for customers in adjustable conditions.

This parallel design enabled direct comparison between human and AI responses since both received equivalent scenario information and responded using identical evaluation criteria and response formats. The standardized vignette approach ensured that any observed differences between human and AI behavior could be attributed to decision-making processes rather than variations in experimental stimuli or response requirements.

\begin{figure}[!htb]
\singlespacing
\begin{lstlisting}
LLM DYADIC BEHAVIOR SIMULATION PROCESS
INPUT: Standardized vignettes, experimental conditions, LLM models
OUTPUT: Customer and driver satisfaction ratings with reasoning
1. SETUP
   - Load standardized vignettes for all experimental conditions
   - Initialize LLM API connections (GPT-4o, GPT-4.1, Sonnet 3.5, Sonnet 4, Mistral Large 2, Mistral Medium 3)
   - Set experimental parameters (30 iterations per condition)
2. FOR each experimental condition combination:
   a. Select tip_adjustable [TRUE/FALSE]
   b. Select tip_visibility ['before'/'after'] 
   c. Select service_outcome ['fails'/'below'/'meets'/'exceeds']
   d. FOR each LLM model:
      e. FOR 30 iterations:
         f. Select matching vignette based on experimental conditions
         g. Create customer prompt:
            - Add vignette text
            - Present scenario outcome text
            - If tip_adjustable: request tip decision and amount
            - Request satisfaction rating (1-7) and reasoning
         h. Create driver prompt:
            - Add matching vignette text
            - Present matching scenario outcome text with customer response
            - Request satisfaction rating (1-7) and reasoning
         i. Call LLM API for customer response
         j. Call LLM API for driver response  
         k. Parse responses to extract:
            - Satisfaction ratings and reasoning
            - Tip decisions and amounts (if applicable)
         l. Store results with experimental metadata
3. SAVE all results to structured dataset for validation analysis
RESPONSE PARSING FUNCTION:
   - Extract satisfaction ratings using regex patterns
   - Extract reasoning text and tip decisions
   - Handle parsing errors and missing data
\end{lstlisting}
\caption{Pseudocode describing the programmatic implementation of the LLM dyadic behavior simulation. Each experimental condition is replicated 30 times per model to generate human-comparable behavioral data.}
\label{alg:llm_simulation_process}
\end{figure}

\FloatBarrier
\clearpage
\section{Games-Howell Pairwise Comparisons} \label{app:GHcomparisons}
This section presents the detailed pairwise comparison results from the Games-Howell post-hoc tests for the three behavioral measures.

\begin{table}[!ht]
\centering
\caption{Games-Howell Post-hoc Comparisons for Tip Change Behavior}
\label{tab:posthoc_tip}
\begin{tabular}{ll*{2}{S[table-format=-1.3]}}
\toprule
\textbf{Model A} & \textbf{Model B} & {\textbf{Difference}} & {\textbf{p-value}} \\
\midrule
GPT-4.1 & GPT-4o & -1.019 & 0.008 \\
GPT-4.1 & Human & -0.790 & 0.101 \\
GPT-4.1 & Mistral Large 2 & -0.940 & 0.033 \\
GPT-4.1 & Mistral Medium 3 & -1.270 & 0.001 \\
GPT-4.1 & Sonnet 3.5 & -0.980 & 0.012 \\
GPT-4.1 & Sonnet 4 & -1.166 & 0.004 \\
GPT-4o & Human & 0.228 & 0.953 \\
GPT-4o & Mistral Large 2 & 0.079 & 1.000 \\
GPT-4o & Mistral Medium 3 & -0.251 & 0.943 \\
GPT-4o & Sonnet 3.5 & 0.039 & 1.000 \\
GPT-4o & Sonnet 4 & -0.147 & 0.997 \\
Human & Mistral Large 2 & -0.149 & 0.997 \\
Human & Mistral Medium 3 & -0.479 & 0.447 \\
Human & Sonnet 3.5 & -0.190 & 0.980 \\
Human & Sonnet 4 & -0.375 & 0.781 \\
Mistral Large 2 & Mistral Medium 3 & -0.330 & 0.859 \\
Mistral Large 2 & Sonnet 3.5 & -0.041 & 1.000 \\
Mistral Large 2 & Sonnet 4 & -0.226 & 0.982 \\
Mistral Medium 3 & Sonnet 3.5 & 0.290 & 0.887 \\
Mistral Medium 3 & Sonnet 4 & 0.104 & 1.000 \\
Sonnet 3.5 & Sonnet 4 & -0.185 & 0.991 \\
\bottomrule
\end{tabular}
\end{table}

\begin{table}[!ht]
\centering
\caption{Games-Howell Post-hoc Comparisons for Joint Satisfaction}
\label{tab:posthoc_joint}
\begin{tabular}{ll*{2}{S[table-format=-1.3]}}
\toprule
\textbf{Model A} & \textbf{Model B} & {\textbf{Difference}} & {\textbf{p-value}} \\
\midrule
GPT-4.1 & GPT-4o & 0.126 & 0.975 \\
GPT-4.1 & Human & 0.122 & 0.974 \\
GPT-4.1 & Mistral Large 2 & 0.243 & 0.580 \\
GPT-4.1 & Mistral Medium 3 & 0.316 & 0.276 \\
GPT-4.1 & Sonnet 3.5 & 0.347 & 0.153 \\
GPT-4.1 & Sonnet 4 & 0.218 & 0.672 \\
GPT-4o & Human & -0.004 & 1.000 \\
GPT-4o & Mistral Large 2 & 0.117 & 0.983 \\
GPT-4o & Mistral Medium 3 & 0.190 & 0.850 \\
GPT-4o & Sonnet 3.5 & 0.221 & 0.711 \\
GPT-4o & Sonnet 4 & 0.092 & 0.995 \\
Human & Mistral Large 2 & 0.121 & 0.975 \\
Human & Mistral Medium 3 & 0.194 & 0.806 \\
Human & Sonnet 3.5 & 0.225 & 0.645 \\
Human & Sonnet 4 & 0.096 & 0.991 \\
Mistral Large 2 & Mistral Medium 3 & 0.073 & 0.999 \\
Mistral Large 2 & Sonnet 3.5 & 0.104 & 0.989 \\
Mistral Large 2 & Sonnet 4 & -0.025 & 1.000 \\
Mistral Medium 3 & Sonnet 3.5 & 0.031 & 1.000 \\
Mistral Medium 3 & Sonnet 4 & -0.098 & 0.992 \\
Sonnet 3.5 & Sonnet 4 & -0.129 & 0.962 \\
\bottomrule
\end{tabular}
\end{table}

\begin{table}[!ht]
\centering
\caption{Games-Howell Post-hoc Comparisons for Differential Satisfaction}
\label{tab:posthoc_diff}
\begin{tabular}{ll*{2}{S[table-format=-1.3]}}
\toprule
\textbf{Model A} & \textbf{Model B} & {\textbf{Difference}} & {\textbf{p-value}} \\
\midrule
GPT-4.1 & GPT-4o & 0.348 & 0.004 \\
GPT-4.1 & Human & 0.399 & 0.014 \\
GPT-4.1 & Mistral Large 2 & -0.360 & 0.001 \\
GPT-4.1 & Mistral Medium 3 & -0.477 & {$<$0.001} \\
GPT-4.1 & Sonnet 3.5 & 0.265 & 0.065 \\
GPT-4.1 & Sonnet 4 & -0.315 & 0.008 \\
GPT-4o & Human & 0.051 & 0.999 \\
GPT-4o & Mistral Large 2 & -0.708 & {$<$0.001} \\
GPT-4o & Mistral Medium 3 & -0.825 & {$<$0.001} \\
GPT-4o & Sonnet 3.5 & -0.083 & 0.966 \\
GPT-4o & Sonnet 4 & -0.663 & {$<$0.001} \\
Human & Mistral Large 2 & -0.760 & {$<$0.001} \\
Human & Mistral Medium 3 & -0.876 & {$<$0.001} \\
Human & Sonnet 3.5 & -0.135 & 0.906 \\
Human & Sonnet 4 & -0.714 & {$<$0.001} \\
Mistral Large 2 & Mistral Medium 3 & -0.117 & 0.741 \\
Mistral Large 2 & Sonnet 3.5 & 0.625 & {$<$0.001} \\
Mistral Large 2 & Sonnet 4 & 0.046 & 0.998 \\
Mistral Medium 3 & Sonnet 3.5 & 0.742 & {$<$0.001} \\
Mistral Medium 3 & Sonnet 4 & 0.162 & 0.368 \\
Sonnet 3.5 & Sonnet 4 & -0.579 & {$<$0.001} \\
\bottomrule
\end{tabular}
\end{table}

\FloatBarrier
\clearpage
\section{Structural Equation Modeling Results} \label{app:SEMresults}
This section presents the results of the multi-group SEM analysis (Tables~\ref{tab:sem_humans}-\ref{tab:sem_mistral_medium}). The indirect effects estimates are provided in Table~\ref{tab:bootstrap_all}.

\begin{table}[!ht]
\centering
\caption{SEM Results - Human Participants Only}
\label{tab:sem_humans}
\begin{tabular}{lll*{3}{S[table-format=-1.3]}}
\toprule
\textbf{LHS} & \textbf{Op} & \textbf{RHS} & {\textbf{Estimate}} & {\textbf{Std. Error}} & {\textbf{p-value}} \\
\midrule
\multicolumn{6}{l}{\textit{Regression Paths}} \\
Tip change & $\sim$ & Service outcome & 1.169 & 0.160 & \textbf{$<$0.001} \\
Tip change & $\sim$ & Service outcome $\times$ Adjustability & 0.104 & 0.163 & 0.526 \\
Joint satisfaction & $\sim$ & Tip change & 0.051 & 0.013 & \textbf{$<$0.001} \\
Joint satisfaction & $\sim$ & Tip change $\times$ Visibility & 0.017 & 0.018 & 0.345 \\
Joint satisfaction & $\sim$ & Service outcome & 1.557 & 0.045 & \textbf{$<$0.001} \\
Differential satisfaction & $\sim$ & Tip change & 0.063 & 0.025 & \textbf{0.012} \\
Differential satisfaction & $\sim$ & Tip change $\times$ Visibility & -0.175 & 0.033 & \textbf{$<$0.001} \\
Differential satisfaction & $\sim$ & Service outcome & 0.936 & 0.085 & \textbf{$<$0.001} \\
\midrule
\multicolumn{6}{l}{\textit{Variances}} \\
Tip change & $\sim\sim$ & Tip change & 11.127 & 0.720 & \textbf{$<$0.001} \\
Differential satisfaction & $\sim\sim$ & Differential satisfaction & 3.359 & 0.217 & \textbf{$<$0.001} \\
Joint satisfaction & $\sim\sim$ & Joint satisfaction & 0.959 & 0.062 & \textbf{$<$0.001} \\
\bottomrule
\end{tabular}
\end{table}

\begin{table}[!ht]
\centering
\caption{SEM Results - GPT-4o Only}
\label{tab:sem_GPT4o}
\begin{tabular}{lll*{3}{S[table-format=-1.3]}}
\toprule
\textbf{LHS} & \textbf{Op} & \textbf{RHS} & {\textbf{Estimate}} & {\textbf{Std. Error}} & {\textbf{p-value}} \\
\midrule
\multicolumn{6}{l}{\textit{Regression Paths}} \\
Tip change & $\sim$ & Service outcome & 1.322 & 0.135 & \textbf{$<$0.001} \\
Tip change & $\sim$ & Service outcome $\times$ Adjustability & 0.593 & 0.139 & \textbf{$<$0.001} \\
Joint satisfaction & $\sim$ & Tip change & 0.023 & 0.011 & \textbf{0.041} \\
Joint satisfaction & $\sim$ & Tip change $\times$ Visibility & 0.031 & 0.014 & \textbf{0.032} \\
Joint satisfaction & $\sim$ & Service outcome & 1.860 & 0.037 & \textbf{$<$0.001} \\
Differential satisfaction & $\sim$ & Tip change & 0.004 & 0.018 & 0.803 \\
Differential satisfaction & $\sim$ & Tip change $\times$ Visibility & -0.046 & 0.022 & \textbf{0.040} \\
Differential satisfaction & $\sim$ & Service outcome & 0.754 & 0.056 & \textbf{$<$0.001} \\
\midrule
\multicolumn{6}{l}{\textit{Variances}} \\
Tip change & $\sim\sim$ & Tip change & 8.113 & 0.524 & \textbf{$<$0.001} \\
Differential satisfaction & $\sim\sim$ & Differential satisfaction & 1.241 & 0.080 & \textbf{$<$0.001} \\
Joint satisfaction & $\sim\sim$ & Joint satisfaction & 0.522 & 0.034 & \textbf{$<$0.001} \\
\bottomrule
\end{tabular}
\end{table}

\begin{table}[!ht]
\centering
\caption{SEM Results - GPT-4.1 Only}
\label{tab:sem_GPT41}
\begin{tabular}{lll*{3}{S[table-format=-1.3]}}
\toprule
\textbf{LHS} & \textbf{Op} & \textbf{RHS} & {\textbf{Estimate}} & {\textbf{Std. Error}} & {\textbf{p-value}} \\
\midrule
\multicolumn{6}{l}{\textit{Regression Paths}} \\
Tip change & $\sim$ & Service outcome & 2.462 & 0.206 & \textbf{$<$0.001} \\
Tip change & $\sim$ & Service outcome $\times$ Adjustability & 0.503 & 0.211 & \textbf{0.017} \\
Joint satisfaction & $\sim$ & Tip change & -0.035 & 0.005 & \textbf{$<$0.001} \\
Joint satisfaction & $\sim$ & Tip change $\times$ Visibility & -0.005 & 0.009 & 0.568 \\
Joint satisfaction & $\sim$ & Service outcome & 1.953 & 0.027 & \textbf{$<$0.001} \\
Differential satisfaction & $\sim$ & Tip change & 0.136 & 0.013 & \textbf{$<$0.001} \\
Differential satisfaction & $\sim$ & Tip change $\times$ Visibility & -0.092 & 0.021 & \textbf{$<$0.001} \\
Differential satisfaction & $\sim$ & Service outcome & 0.349 & 0.063 & \textbf{$<$0.001} \\
\midrule
\multicolumn{6}{l}{\textit{Variances}} \\
Tip change & $\sim\sim$ & Tip change & 18.770 & 1.212 & \textbf{$<$0.001} \\
Differential satisfaction & $\sim\sim$ & Differential satisfaction & 1.432 & 0.092 & \textbf{$<$0.001} \\
Joint satisfaction & $\sim\sim$ & Joint satisfaction & 0.260 & 0.017 & \textbf{$<$0.001} \\
\bottomrule
\end{tabular}
\end{table}

\begin{table}[!ht]
\centering
\caption{SEM Results - Sonnet-3.5 Only}
\label{tab:sem_sonnet35}
\begin{tabular}{lll*{3}{S[table-format=-1.3]}}
\toprule
\textbf{LHS} & \textbf{Op} & \textbf{RHS} & {\textbf{Estimate}} & {\textbf{Std. Error}} & {\textbf{p-value}} \\
\midrule
\multicolumn{6}{l}{\textit{Regression Paths}} \\
Tip change & $\sim$ & Service outcome & 1.217 & 0.138 & \textbf{$<$0.001} \\
Tip change & $\sim$ & Service outcome $\times$ Adjustability & 0.426 & 0.142 & \textbf{0.003} \\
Joint satisfaction & $\sim$ & Tip change & 0.021 & 0.010 & \textbf{0.047} \\
Joint satisfaction & $\sim$ & Tip change $\times$ Visibility & 0.017 & 0.013 & 0.201 \\
Joint satisfaction & $\sim$ & Service outcome & 1.755 & 0.033 & \textbf{$<$0.001} \\
Differential satisfaction & $\sim$ & Tip change & -0.036 & 0.015 & \textbf{0.020} \\
Differential satisfaction & $\sim$ & Tip change $\times$ Visibility & -0.064 & 0.020 & \textbf{0.001} \\
Differential satisfaction & $\sim$ & Service outcome & 0.916 & 0.048 & \textbf{$<$0.001} \\
\midrule
\multicolumn{6}{l}{\textit{Variances}} \\
Tip change & $\sim\sim$ & Tip change & 8.422 & 0.544 & \textbf{$<$0.001} \\
Differential satisfaction & $\sim\sim$ & Differential satisfaction & 0.988 & 0.064 & \textbf{$<$0.001} \\
Joint satisfaction & $\sim\sim$ & Joint satisfaction & 0.452 & 0.029 & \textbf{$<$0.001} \\
\bottomrule
\end{tabular}
\end{table}

\begin{table}[!ht]
\centering
\caption{SEM Results - Sonnet-4 Only}
\label{tab:sem_sonnet4}
\begin{tabular}{lll*{3}{S[table-format=-1.3]}}
\toprule
\textbf{LHS} & \textbf{Op} & \textbf{RHS} & {\textbf{Estimate}} & {\textbf{Std. Error}} & {\textbf{p-value}} \\
\midrule
\multicolumn{6}{l}{\textit{Regression Paths}} \\
Tip change & $\sim$ & Service outcome & 1.376 & 0.200 & \textbf{$<$0.001} \\
Tip change & $\sim$ & Service outcome $\times$ Adjustability & 0.430 & 0.185 & \textbf{0.020} \\
Joint satisfaction & $\sim$ & Tip change & 0.038 & 0.006 & \textbf{$<$0.001} \\
Joint satisfaction & $\sim$ & Tip change $\times$ Visibility & -0.009 & 0.008 & 0.255 \\
Joint satisfaction & $\sim$ & Service outcome & 1.699 & 0.025 & \textbf{$<$0.001} \\
Differential satisfaction & $\sim$ & Tip change & 0.136 & 0.011 & \textbf{$<$0.001} \\
Differential satisfaction & $\sim$ & Tip change $\times$ Visibility & -0.140 & 0.013 & \textbf{$<$0.001} \\
Differential satisfaction & $\sim$ & Service outcome & 0.692 & 0.044 & \textbf{$<$0.001} \\
\midrule
\multicolumn{6}{l}{\textit{Variances}} \\
Tip change & $\sim\sim$ & Tip change & 14.369 & 0.987 & \textbf{$<$0.001} \\
Differential satisfaction & $\sim\sim$ & Differential satisfaction & 0.687 & 0.047 & \textbf{$<$0.001} \\
Joint satisfaction & $\sim\sim$ & Joint satisfaction & 0.232 & 0.016 & \textbf{$<$0.001} \\
\bottomrule
\end{tabular}
\end{table}

\begin{table}[!ht]
\centering
\caption{SEM Results - Mistral-Large-2 Only}
\label{tab:sem_mistral_large}
\begin{tabular}{lll*{3}{S[table-format=-1.3]}}
\toprule
\textbf{LHS} & \textbf{Op} & \textbf{RHS} & {\textbf{Estimate}} & {\textbf{Std. Error}} & {\textbf{p-value}} \\
\midrule
\multicolumn{6}{l}{\textit{Regression Paths}} \\
Tip change & $\sim$ & Service outcome & 1.511 & 0.158 & \textbf{$<$0.001} \\
Tip change & $\sim$ & Service outcome $\times$ Adjustability & 0.698 & 0.162 & \textbf{$<$0.001} \\
Joint satisfaction & $\sim$ & Tip change & 0.013 & 0.010 & 0.184 \\
Joint satisfaction & $\sim$ & Tip change $\times$ Visibility & 0.014 & 0.013 & 0.256 \\
Joint satisfaction & $\sim$ & Service outcome & 1.771 & 0.036 & \textbf{$<$0.001} \\
Differential satisfaction & $\sim$ & Tip change & -0.007 & 0.015 & 0.623 \\
Differential satisfaction & $\sim$ & Tip change $\times$ Visibility & -0.021 & 0.019 & 0.280 \\
Differential satisfaction & $\sim$ & Service outcome & 0.516 & 0.056 & \textbf{$<$0.001} \\
\midrule
\multicolumn{6}{l}{\textit{Variances}} \\
Tip change & $\sim\sim$ & Tip change & 11.038 & 0.713 & \textbf{$<$0.001} \\
Differential satisfaction & $\sim\sim$ & Differential satisfaction & 1.220 & 0.079 & \textbf{$<$0.001} \\
Joint satisfaction & $\sim\sim$ & Joint satisfaction & 0.508 & 0.033 & \textbf{$<$0.001} \\
\bottomrule
\end{tabular}
\end{table}

\begin{table}[!ht]
\centering
\caption{SEM Results - Mistral-Medium-3 Only}
\label{tab:sem_mistral_medium}
\begin{tabular}{lll*{3}{S[table-format=-1.3]}}
\toprule
\textbf{LHS} & \textbf{Op} & \textbf{RHS} & {\textbf{Estimate}} & {\textbf{Std. Error}} & {\textbf{p-value}} \\
\midrule
\multicolumn{6}{l}{\textit{Regression Paths}} \\
Tip change & $\sim$ & Service outcome & 1.245 & 0.179 & \textbf{$<$0.001} \\
Tip change & $\sim$ & Service outcome $\times$ Adjustability & 0.441 & 0.165 & \textbf{0.008} \\
Joint satisfaction & $\sim$ & Tip change & 0.033 & 0.007 & \textbf{$<$0.001} \\
Joint satisfaction & $\sim$ & Tip change $\times$ Visibility & -0.017 & 0.009 & 0.064 \\
Joint satisfaction & $\sim$ & Service outcome & 1.963 & 0.027 & \textbf{$<$0.001} \\
Differential satisfaction & $\sim$ & Tip change & 0.137 & 0.013 & \textbf{$<$0.001} \\
Differential satisfaction & $\sim$ & Tip change $\times$ Visibility & -0.103 & 0.017 & \textbf{$<$0.001} \\
Differential satisfaction & $\sim$ & Service outcome & 0.196 & 0.050 & \textbf{$<$0.001} \\
\midrule
\multicolumn{6}{l}{\textit{Variances}} \\
Tip change & $\sim\sim$ & Tip change & 11.455 & 0.789 & \textbf{$<$0.001} \\
Differential satisfaction & $\sim\sim$ & Differential satisfaction & 0.890 & 0.061 & \textbf{$<$0.001} \\
Joint satisfaction & $\sim\sim$ & Joint satisfaction & 0.248 & 0.017 & \textbf{$<$0.001} \\
\bottomrule
\end{tabular}
\end{table}

\begin{sidewaystable}[!ht]
\centering
\begin{threeparttable}
\caption{Bootstrapped Indirect Effects Comparison Across All Groups}
\label{tab:bootstrap_all}
\begin{tabular}{ll*{5}{S[table-format=-1.4]}l}
\toprule
\textbf{Group} & \textbf{Effect} & {\textbf{Estimate}} & {\textbf{Bootstrap SE}} & {\textbf{CI 2.5\%}} & {\textbf{CI 97.5\%}} & {\textbf{p-value}} & \textbf{Significant} \\
\midrule
Human & Indirect\_joint & 0.002 & 0.0043 & -0.0043 & 0.0134 & 0.6064 & No \\
Human & Indirect\_diff & -0.018 & 0.0179 & -0.0558 & 0.0158 & 0.2868 & No \\
GPT-4o & Indirect\_joint & 0.018 & 0.0093 & 0.0014 & 0.0382 & \textbf{0.0368} & Yes \\
GPT-4o & Indirect\_diff & -0.027 & 0.0123 & -0.0521 & -0.0028 & \textbf{0.0332} & Yes \\
GPT-4.1 & Indirect\_joint & -0.003 & 0.0030 & -0.0079 & 0.0038 & 0.3948 & No \\
GPT-4.1 & Indirect\_diff & -0.046 & 0.0134 & -0.0755 & -0.0235 & \textbf{$<$0.001} & Yes \\
Sonnet 3.5 & Indirect\_joint & 0.007 & 0.0056 & -0.0042 & 0.0183 & 0.1976 & No \\
Sonnet 3.5 & Indirect\_diff & -0.027 & 0.0091 & -0.0448 & -0.0092 & \textbf{0.0024} & Yes \\
Sonnet 4 & Indirect\_joint & -0.004 & 0.0032 & -0.0106 & 0.0019 & 0.1988 & No \\
Sonnet 4 & Indirect\_diff & -0.060 & 0.0229 & -0.1096 & -0.0221 & \textbf{0.0008} & Yes \\
Mistral Large 2 & Indirect\_joint & 0.010 & 0.0104 & -0.0085 & 0.0320 & 0.2988 & No \\
Mistral Large 2 & Indirect\_diff & -0.015 & 0.0139 & -0.0425 & 0.0126 & 0.2784 & No \\
Mistral Medium 3 & Indirect\_joint & -0.007 & 0.0052 & -0.0183 & 0.0021 & 0.1336 & No \\
Mistral Medium 3 & Indirect\_diff & -0.045 & 0.0198 & -0.0914 & -0.0148 & \textbf{$<$0.001} & Yes \\
\bottomrule
\end{tabular}
\begin{tablenotes}
\small
\item \textit{Note:} Bootstrap samples = 5,000. Confidence intervals based on bias-corrected percentile method. Significant effects (p < 0.05) are bolded.
\end{tablenotes}
\end{threeparttable}
\end{sidewaystable}

\end{document}